# Great expectations: Unifying Statistical Theory and Programming

Bradley Saul
bradleysaul@fastmail.com

**Abstract** Beginning in the 1970s, statistician-cum-logician Per Martin-Löf wrote a series of papers developing what became Martin-Löf type theory, realizing a system where the distinction between mathematics and programming disappears. Inspired by this vision, this paper introduces dependent type theory (of which Martin-Löf type theory is an example) to a statistical audience. Examples from statistics and probability theory demonstrate how dependent type theory and an algebraic perspective can unify the theoretical and computational concerns of statistics, ensuring rigorous, machine-checked proofs and executable software.

## 1 Introduction

Take nothing on its looks; take everything on evidence. There's no better rule.

— from Great Expectations by Charles Dickens

Statistical methods are traditionally developed in two distinct but related disciplines: mathematics (theory) and programming (implementation). On the theoretical side, statisticians derive estimators, prove properties, and establish asymptotic results using probability theory and linear algebra. On the computational side, programmers implement these methods in software, often relying on numerical libraries and optimization routines. While this separation of concerns has historically been practical, it introduces significant risks: mathematical proofs are often informal, incomplete, or "left to the reader," making them prone to human error; implementations are rarely formally verified against the theoretical specifications, leaving room for mismatches between theory and practice; and the lack of a unified framework makes it difficult to ensure that statistical methods are both mathematically and computationally correct. These issues undermine the reliability and reproducibility of statistical results and scientific software more broadly (Black 2019; Gokhale et al. 2023; Hill et al. 2024; Keeling and Pavur 2007; Mccullough 1998; 2000), with far-reaching consequences for science and decision-making.

The consequences of the theory-implementation gap in statistics are not merely academic. High-profile examples include reproducibility failures in cancer research where statistical software implementations did not match their theoretical descriptions (Baggerly and Coombes 2009), and economic policy decisions based on spreadsheet coding errors (Herndon et al. 2014). Commercially developed statistical software has been found to produce different results for identical analyses (McCullough and Heiser 2008; Yalta and Jenal 2009) and open-source statistical packages have implemented methods that deviate from their mathematical specifications (Almiron et al. 2010; Black 2019; Dominici et al. 2002; Horton and Kleinman 2007; Zeileis and Kleiber 2005), highlighting the need for frameworks that guarantee alignment between statistical theory and implementation.

A solution to this problem should integrate logical and computational concerns. Methodologists should be able to "do math" in a way that allows their proofs to be machine-checked, ensuring rigor and eliminating human error. This mathematical framework should not only support formal reasoning but also compile directly to executable implementations, bridging the gap between theory and practice. Implementations must include proof of faithfulness to their mathematical specifications – a guarantee that is often missing in traditional approaches. In this paper, dependent type theory (Bove and Dybjer 2008; Dybjer and Palmgren 2024) is presented as a solution, demonstrating how it unifies logical and computational content, enables formal verification, and provides a foundation for both mathematical reasoning and statistical software. This paper is intended for statisticians interested in verifiable modeling frameworks, where probability, estimation, and inference are expressed in a formal language that serves both as mathematics and as code. This allows statistical reasoning to be stated precisely and checked mechanically.

Dependent type theory provides a foundation for both mathematical reasoning and computation. Unlike traditional approaches where mathematical theorems and their implementations are separate artifacts, Dependent type theory treats proofs as executable programs. At the heart of dependent type theory is the Curry-Howard correspondence: the idea that types are propositions and programs are proofs. The Curry-Howard correspondence (Dybjer and Palmgren 2024; Sørensen and Urzyczyn 2006; Wadler 2015) establishes a deep connection between logic and programming: types correspond to propositions and programs correspond to proofs. This correspondence allows us to express mathematical theorems as types. For example, the type representing "a specific estimator is unbiased for a given parameter" would correspond to a mathematical theorem, and a program of that type would constitute a proof of that theorem.

The goal of correct-by-construction software is neither new nor unique to statistics, tracing back to the "software crisis" identified in the late 1960s (Naur and Randell 1969), when growing program complexity began outpacing developers' ability to ensure reliability. Formal methods emerged as a subdiscipline of computer science dedicated to the mathematically rigorous specification and verification of software systems. While these approaches initially remained largely theoretical or restricted to critical domains like aerospace and nuclear control, the past two decades have seen a remarkable expansion in programming languages that combine logical expressiveness with computational power. The development of interactive theorem provers such as Coq (Team 2024), Lean (Moura and Ullrich 2021), Isabelle/HOL, and Agda (Agda Developers 2024) has made formal verification increasingly accessible to practitioners beyond specialized niches. This convergence of logical foundations and practical programming now brings the long-sought goal of correct-by-construction software within reach for statistical applications. It offers a principled path to bridge the gap between mathematical rigor and computational implementation.

Using the programming language Agda (Agda Developers 2024), this paper introduces dependent type theory to a statistical audience by formalizing elementary statistical concepts, proving foundational theorems, and solving an example statistical problem. The approach herein offers several advantages beyond correctness guarantees. First, it enables safer evolution of statistical methods, as changes to algorithms require corresponding updates to proofs. Second, it supports modular reasoning, allowing statisticians to build on verified components without needing to understand all implementation details. Finally, this approach bridges disciplinary boundaries, giving mathematically-inclined statisticians and computationally-focused programmers a common language for collaboration.

This paper presents statistical theory as in (Harremoës 2025; Whittle 2012), where expectation is taken as the primitive notion. This perspective parallels the measure-theoretic view in which a probability measure defines a positive, normalized linear functional on random variables, but is developed here in a constructive form so that theorems and implementations share the same formal basis. Broader questions about how this framework relates to other probabilistic foundations lie outside the present scope.

The rest of the paper is organized as follows. Section 2 provides a brief introduction to dependent type theory before presenting three preliminary examples. Section 3 defines the expectation type and derived concepts such as probability and conditional random variables. Then Section 4 introduces the primitive operations for constructing and combining expectation types, and several useful derived operations such as convolution are shown. In Section 5, the expectation type and the operations from Section 4 are put to work to construct several examples of expectation spaces, including several "named" distributions. Some examples include numerical computations on rational numbers to show correct computations fall out of the type without explicitly defining a mass or density function. Finally, Section 6 discusses the implications of dependent type theory, its limitations, existing work in this area, and directions for future work.

# 2 Setting Expectations: Brief Introduction to Dependent Type Theory

Type theory, like set theory, is a foundational system used to reason about mathematical objects. While set theory builds mathematics from the concept of sets and their membership relations, type theory starts from the concept of rules for how objects are constructed and used.

Dependent type theory extends simple type theory by allowing types to depend on values. This extension enables precise specifications of mathematical structures and their properties, creating a formal system as powerful as first-order logic. This paper uses Agda, a programming language that implements constructive type theory. In this framework, formal proofs provide explicit computational content: to give a proof is to give an algorithm.

## 2.1 From Simple Types to Dependent Types

Conventional programming languages with static typing have types like `Bool` for true/false values, `Double` for numerical values, or `List` for collections. However, many statistical concepts involve constraints that simple types cannot express; e.g.:

- A correlation matrix must be square and symmetric
- A probability must be between 0 and 1
- A sample size must be positive

Dependent types allow us to encode these constraints directly in the type system.

## 2.2 Application to Statistics

Statisticians often want to prove properties like:

- an estimator is unbiased;
- a statistical test maintains its significance level;
- or a numerical algorithm converges.

In a programming language with dependent types, one can both express such properties as types and write proofs of these properties, which can be verified by the language's type-checker. This paper is a literate Agda program. All code in this paper has been type-checked by Agda, though some code has been elided from the type-set document for clarity. Type checking ensures that all definitions and proofs are logically valid with respect to the underlying formal system, ruling out inconsistencies, type mismatches, and unproven claims. The complete code base can be found at the project website. An introduction to Agda's syntax is provided in Appendix A.

The three examples that follow demonstrate statistical concepts expressed in Agda. The types used in the examples, such as the expectation family of types ($\mathcal{E}$), are defined in later sections. Readers should focus on the statistical meaning and the type-level guarantees, not the syntax. Details will be explained in later sections. The examples include two classical theorems, the Law of the Unconscious Statistician and the tower property for conditional expectation, and an example distribution, the hypergeometric law.

## 2.3 Example: Law of Unconscious Statistician

The Law of the Unconscious Statistician (LOTUS) states that the expectation of a random variable composed with a transformation equals the expectation of the original random variable under the transformed expectation space. In the framework presented in this paper, LOTUS follows immediately from the definitions:

```
module LOTUS
  (S T : SampleSpace) -- Given 2 sample spaces
  (X : T → Mₛ)        -- Random variable on T
  (f : S → T)         -- map from S to T
  where

  open ℰ

  theorem : ∀ {𝔼 : ℰ S} → E 𝔼 (X ∘→ f) ≈ᴹ E (𝔼 ⟨$⟩ f) X
  theorem = ≈ᴹ-refl
```

The proof is reflexivity. This demonstrates that LOTUS is a definitional consequence of how expectation and the pushforward measure (here, the `⟨$⟩` operation) interact.

## 2.4 Example: Law of Total Expectation

The conditional expectation `E[Y|X]` is characterized by an orthogonality condition: the residual `Y − E[Y|X]` has zero expectation against all functions of `X`. This makes `E[Y|X]` the best predictor of `Y` given `X`. From this property the law of total expectation, also called the tower property, follows:

```
module _ {Y X} (𝔼ₓ : E[ Y | X ]) where
  open E[_|_] 𝔼ₓ

  tower : E E[Y|X] ≈ᴹ E Y
  tower = begin
    E E[Y|X]                  ≈⟨ E-cong (+ᴹ-identityʳ _) ⟩
    E (E[Y|X] +′ 0ᴹ′)         ≈⟨ E-cong (+ᴹ-congˡ (-ᴹ‿inverseˡ _)) ⟩
    E ((E[Y|X] +′ (-′ Y +′ Y))) ≈⟨ E-cong (+ᴹ-assoc _ _ _) ⟩
    E (E[Y|X] -′ Y +′ Y)      ≈⟨ +-homo ⟩
    E (E[Y|X] -′ Y) +ᴹ E Y    ≈⟨ +ᴹ-congʳ E[EY|X-Y]≈0 ⟩
    0ᴹ +ᴹ E Y                 ≈⟨ +ᴹ-identityˡ (E Y) ⟩
    E Y                       ∎
```

The proof is written in equational reasoning style. The chain of steps begins with the left-hand side of the theorem and ends with the right-hand side, so the entire derivation reads as a calculation in which every step is justified by a named theorem. See Section 3.3 for an example that explains each reasoning step in more detail.

## 2.5 Example: Hypergeometric Distribution

Here, the construction of the hypergeometric probability distribution is shown. The `urnTrials` (Section 5.2.2.1) function models the changing composition of the urn after each draw, ensuring the correct dependency between trials. The `countUptoK` function then transforms each sequence of outcomes into the number of successes observed in `n` draws, yielding the hypergeometric distribution over $\{0, 1, \ldots, K\}$.

```
hypergeometric : (N K : ℕ) (n : ℕ⁺) → ℰ (Finₛ (K +1))
hypergeometric N K n = urnTrials K (N ∸ K) n ⟨$⟩ countUptoK K n
```

As a concrete case, drawing three times from a population of five items with two marked “success” produces:

```
𝔼 : ℰ (Fin⁺ₛ (3 #))
𝔼 = hypergeometric 5 2 (3 #)
```

```
_ : PR (δF 1F) ≟ + 6 / 10 ; _ =  ✓
```

Here the probability of exactly one success is $\frac{6}{10}$, in agreement with the classical formula. This illustrates how a familiar distribution can be constructed and its properties verified directly inside dependent type theory.

These previews demonstrate how standard statistical laws and familiar distributions can be captured as types and proved directly in dependent type theory. The point is not the novelty of the results themselves but the fact that they emerge naturally once expectation, conditioning, and sampling are given precise type-theoretic form. Section 3 develops the underlying framework in full generality, introducing the structures that make these proofs possible. Additional worked examples of concrete distributions, appear later in Section 5.

# 3 Creating Expectations: Defining the $\mathcal{E}$ Type

This section formalizes the concept of expectation using dependent types. Rather than defining probability first and then deriving expectation, expectation is taken as the primitive concept, following Whittle (2012). Traditional probability theory establishes probability measures on sample spaces, defines random variables as measurable functions, and derives expectation through integration with respect to the probability measure. The expectation-first approach defines expectation operators with required axiomatic properties, derives probability as expectation of indicator functions, and builds statistical theory from expectation spaces’ algebraic structure.

## 3.1 Mathematical Requirements

The formalization of expectation spaces requires three mathematical components that work together to provide a foundation for statistical reasoning. The base algebraic structure is a totally ordered field (`ToField`) denoted $\mathcal{R}$ with carrier type `R`, which provides the scalar arithmetic necessary for statistical computation. The `ToField` requirement ensures that only types with verified field properties can be used, precluding the direct use of floating-point types without explicitly postulating that they satisfy field axioms. Beyond basic field operations (addition, multiplication, and their inverses), the structure includes a total ordering that enables comparison of any two elements, supporting inequality-based reasoning needed for optimization and bounding expectations and probabilities. A `ToField` also includes an apartness relation, provides a constructive alternative to the classical notion of inequality.

An algebra over the field (`MAlgebra`), denoted $\mathcal{M}$ with carrier type `M`, generalizes the codomain of random variables beyond scalar-valued functions. In some applications, scalars are all that is needed, and by the tensor unit construction, any `ToField` can be made into an `MAlgebra` (see Appendix C.2.1). However, many statistical applications involve vector-valued random variables (multivariate analysis), matrix-valued random variables (random matrix theory), or function-valued random variables (functional data analysis). The `MAlgebra` provides the algebraic structure necessary to handle these cases uniformly. For additional details, see Appendix C.

Sample spaces are formalized as setoids, which are types equipped with equivalence relations (see Appendix B for more details). This approach addresses a fundamental issue in statistical theory: when should two outcomes in an experiment be considered equivalent? In discrete probability, the equivalence relation is typically equality. In continuous probability, outcomes that differ only on sets of measure zero should be considered equivalent for statistical purposes. Random variables are required to respect the sample space equivalence relation, ensuring that equivalent outcomes yield equivalent measurement values. See Appendix D for operations, constructions, and properties of random variables. The setoid structure also enables the construction of quotient, product, and co-product spaces with well-defined equivalence relations, supporting complex sample space constructions needed for hierarchical models and structured data analysis.

## 3.2 Formal Definition of Expectation Spaces

The expectation type $\mathcal{E}$ is defined as an Agda `record`, which generalizes dependent products (Σ-types) by providing named fields rather than unnamed projections. The code in the following line declares the construction of a `record` type, named $\mathcal{E}$, which takes a `SampleSpace` as a parameter. The $\mathcal{E}$ type has type `Type₁` because it contains functions that map from types to types, requiring a higher level to avoid inconsistencies in the type system. The subsequent code after the `where` clause defined in subsections below are named fields within the $\mathcal{E}$ type.

```
record ℰ (S : SampleSpace) : Type₁ where
```

### 3.2.1 Expectation Operator: E

The core of the expectation type is the expectation operator, here denoted `E`, whose domain is random variables (functions from `S` to the sample space of the MAlgebra (`Mₛ`)) and whose codomain is `M`. Measurability or other structural constraints on random variables can be added when needed, but are not assumed a priori.

```
    E : (S → Mₛ) → M -- i.e. RVₛ → Mₛ
```

### 3.2.2 Properties of E

1. **Congruence**

The `E` function must satisfy several properties. First, equivalent random variables can be exchanged under application of `E`; i.e., `E` is a congruence. The proposition `E-cong` is described in more familiar, informal language in the comments.

```
    E-cong :
        ∀ {X₁ X₂ : RV}  -- For all random variables X₁ and X₂,
      → X₁ ≈′ X₂         -- if X₁ is (pointwise) equivalent to X₂
      → E X₁ ≈ᴹ E X₂     -- then E X₁ is equivalent to E X₂.
```

The other properties of `E` below can be read in a similar manner.

2. **Linearity**

Further, E is a linear function. It must be homomorphic with respect to addition and scaling:

```
+-homo  : ∀ {X₁ X₂} → E (X₁ +′ X₂) ≈ᴹ E X₁ +ᴹ E X₂
*ₗ-homo : ∀ {c} {X} → c *ₗ E X ≈ᴹ E (c *ₗ′ X)
```

3. **Positivity**

In addition to linearity, (Whittle 2012) specifies other conditions that E must satisfy. The positivity axiom states that if a random variable is non-negative everywhere, then its expectation must also be non-negative:

```
positive : ∀ {X} → 0ᴹ′ ≤′ X → 0ᴹ ≤ᴹ E X
```

4. **Normalization**

The normalization axiom ensures that the expectation of the function constant at 1ᴹ, where 1ᴹ is the multiplicative identity in the MAlgebra, is equal to 1ᴹ:

```
unital : E 1ᴹ′ ≈ᴹ 1ᴹ
```

5. **Commuting Limits**

This property expresses that E is compatible with limits of increasing sequences of functions (monotone convergence):

```
E-lim-comm : ∀ {Xₙ : Sequence RV} {X : RV}
  → Monotonic↗ Xₙ
  → Xₙ ConvergesPointwiseTo X
  → (E ∘ Xₙ) ConvergesTo E X
```

The properties above capture the essential characteristics of expectation per the axioms outlined in Whittle (2012).

6. **Constant Multiplication Compatibility**

Lastly, the following two properties enable multiplication by constant random variables (denoted K) to be factored out of E. These properties are distinct from and not implied by the scalar multiplication property (*ₗ-homo), which allow scalars to commute with respect to E.

```
*ᴹ-const-homoˡ : ∀ {m} {X} → E (K m *′ X) ≈ᴹ m *ᴹ E X
*ᴹ-const-homoʳ : ∀ {m} {X} → E (X *′ K m) ≈ᴹ E X *ᴹ m
```

A note on notation is important here. Throughout this paper, $\mathcal{E}$ denotes the expectation type; it takes a sample space S as a parameter. A specific instance of $\mathcal{E}$ S (an expectation over the sample space S), is typically denoted as $\mathbb{E}$. Specific $\mathbb{E}$ : $\mathcal{E}$ S are sometimes referred to as a "distribution", which is appropriate since $\mathcal{E}$ S fully characterizes a probability distribution through its expectation operator. The symbol E refers to the expectation operator function within a specific $\mathbb{E}$, while Eᶠ is the E function paired with its proof of congruence (E-cong). In summary, $\mathcal{E}$ is the type of expectation spaces, $\mathbb{E}$ is an instance of that type for a specific sample space, and E is the function within an $\mathbb{E}$ that computes expected values of random variables.

## 3.3 Basic Definitions and Theorems of $\mathcal{E}$ Spaces

With the $\mathcal{E}$ type defined, operations that use E can be defined. For example, an important random variable transformation is subtracting its expectation, or centering:

```
_-EX : RV → RV
X -EX = X -′ K (E X)
```

The product moment is the expectation of the product of two random variables:

```
⟨_,_⟩ : RV → RV → M
⟨ X₁ , X₂ ⟩ = E (X₁ *′ X₂)
```

Covariance and variance are defined as `Cov` and `Var`, respectively:

```
Cov : RV → RV → M
Cov X₁ X₂ = ⟨ X₁ -EX , X₂ -EX ⟩

Var : RV → M
Var X = Cov X X
```

In order to prove that the primitive operations for combining $\mathcal{E}$ spaces in Section 4 satisfy the properties of the $\mathcal{E}$ type, some basic lemmas are needed. First, the `positive` axiom implies monotonicity, stated as:

```
E-mono : ∀ {X₁} {X₂} → X₁ ≤′ X₂ → E X₁ ≤ᴹ E X₂
```

While this property can be proven from first principles, all positive linear maps share this property, so first it is shown that `E` is a positive linear map (⊸⁺) between the partially ordered `LeftSemimodule` of random variables and the partially ordered `LeftSemimodule` within the `MAlgebra` $\mathcal{M}$.

```
E-posLinearMap : RV.poLeftSemimodule ⊸⁺ 𝓜.poLeftSemimodule
E-posLinearMap = mk⊸⁺ E E-cong +-homo *ₗ-homo positive
```

Then `E-mono` is a case of the `⊸⁺-mono` theorem for positive linear maps between partially vector spaces over a commutative ring.

```
E-mono = ⊸⁺-Props.⊸⁺-mono RV.poModule poModule E-posLinearMap
```

Similarly, subtraction commutes with respect to `E`, which is true for any linear map between `LeftModules`:

```
-ᴹ‿homo : ∀ {X} → E (-′ X) ≈ᴹ -ᴹ E X
-ᴹ‿homo = ⊸-Props.-ᴹ‿homo RV.leftModule leftModule E-linearMap
```

The `E-homo` and `-ᴹ‿homo` theorems were instantiated from a pre-existing Agda linear algebra library. The following theorems demonstrate Agda's syntactic support for equational, transitive reasoning. Below, `Agda` comments (`--`) are used to explain the code for the `+ᴹ-ᴹ‿homo` theorem in detail.

```
-- Statement of the theorem
+ᴹ-ᴹ‿homo : ∀ {X₁ X₂} → E (X₁ -′ X₂) ≈ᴹ E X₁ -ᴹ E X₂
-- Proof of the theorem
+ᴹ-ᴹ‿homo {X₁} {X₂} =    -- Bring the implicit arguments X₁ and X₂ into scope
  begin                  -- begin the reasoning
    E (X₁ -′ X₂)
  ≈⟨ +-homo ⟩            -- apply +-homo property of E
    E X₁ +ᴹ E (-′ X₂)
```

```
    ≈⟨ +ᴹ-congˡ -ᴹ‿homo ⟩ -- apply +ᴹ-congˡ -ᴹ‿homo
      E X₁ -ᴹ E X₂
    ∎                     -- ∎ = Q.E.D
    -- Bring `begin ... ≈⟨ ... ⟩ ... ∎ syntax into scope:
    where open ≈-Reasoning ≈ᴹ-setoid
```

The `Em≈m` lemma states that expectation of a constant random variable `K` applied to `m` is equivalent to the constant `m`, which follows from the properties of the `MAlgebra` and $\mathcal{E}$.

```
  Em≈m : ∀ m → E (K m) ≈ᴹ m
  Em≈m m = begin
    E (K m)          ≈⟨ E-cong (*ᴹ-identityʳ m) ⟨
    E (K m *′ K 1ᴹ) ≈⟨ *ᴹ-const-homoˡ ⟩
    m *ᴹ E (K 1ᴹ)    ≈⟨ *ᴹ-congˡ unital ⟩
    m *ᴹ 1ᴹ          ≈⟨ *ᴹ-identityʳ m ⟩
    m                ∎
    where open ≈-Reasoning ≈ᴹ-setoid
```

## 3.4 Probability

The probability function (`Pr`) is defined as the trace of expectation applied to indicator functions. An indicator function is a random variable that carries a proof that its range is either the zero or one element in the `MAlgebra` (`0ᴹ` or `1ᴹ`). The `Pr` function first extracts the random variable from the Indicator (`toRV`), applies the expectation operator `E` to get the expected value, then applies the trace operation `Tr` to extract a scalar from the measurement space, and packages the result as a proportion $P$ (a `ToField` value with a proof of being between 0 and 1 inclusive).

```
Pr : Indicator S → 𝑃
Pr X = mk
  ( (Tr ∘ E ∘ toRV) X)
  ( -- proof 0# ≤ Pr X
    (Tr-positive (positive (0≤𝕀 {f = X} (<ᴹ⇒≤ᴹ <ᴹ-nonTrivial))))
  , -- proof Pr X ≤ 1#
    (begin
      (Tr ∘ E ∘ toRV) X ≡⟨⟩
      Tr (E (toRV X))   ≤⟨ Tr-mono (E-mono (𝕀≤1 {f = X} (<ᴹ⇒≤ᴹ <ᴹ-nonTrivial))) ⟩
      Tr (E (K 1ᴹ) )    ≈⟨ Tr-cong unital ⟩
      Tr 1ᴹ             ≈⟨ Tr-unital ⟩
      1#                ∎)
  )
  where open ≤-Reasoning ℛ.poset
```

For convenience, the function `PR = 𝑃.p ∘ Pr` is provided which extract the `R` value from the probability value, factoring out the proof that the value is between 0 and 1 (inclusive).

#### 3.4.0.1 Properties of `Pr`/`PR`

The definition of `Pr` ensures that probabilities are always values in $[0, 1]$ by construction. The lemmas below recover standard probability identities. Proofs follow from the axioms of $\mathcal{E}$ and the algebra of indicator functions; some are elided here for clarity but are fully checked in the accompanying code.

The probability of the entire sample space is 1:

```
normalization : PR ⊤ ℛ.≈ 1#   -- ⊤ is the indicator that always returns 1.
normalization = begin
    Tr (E (K 1ᴹ)) ≈⟨ Tr-cong unital ⟩
    Tr 1ᴹ         ≈⟨ Tr-unital ⟩
    1#            ∎
    where open ≈-Reasoning ℛ.setoid
```

The probabilities of an event (an indicator function) and its complement sum to one, and their expectations decompose accordingly.

```
  totalE : E X +ᴹ E Xᶜ ≈ᴹ 1ᴹ
  totalE = begin
    E X +ᴹ E Xᶜ ≈⟨ +-homo ⟨
    E (X +′ Xᶜ) ≈⟨ E-cong (I+CI≈1 _ I) ⟩
    E (K 1ᴹ)    ≈⟨ unital ⟩
    1ᴹ          ∎
    where open ≈-Reasoning ≈ᴹ-setoid

  totalProb : PR I ℛ.+ PR (C I) ℛ.≈ 1#
  totalProb = begin
    PR I ℛ.+ PR (C I)      ≡⟨⟩
    Tr (E X) ℛ.+ Tr (E Xᶜ) ≈⟨ Tr-+-homo ⟨
    Tr (E X +ᴹ E Xᶜ)       ≈⟨ Tr-cong totalE ⟩
    Tr 1ᴹ                  ≈⟨ Tr-unital ⟩
    1#                     ∎
    where open ≈-Reasoning ℛ.setoid
```

Here `X` is the random variable corresponding to indicator `I`, and `Xᶜ` is the random variable for the complement indicator. The `totalE` lemma states the expectation form of complementarity; `totalProb` is its scalar-valued consequence for probabilities.

An indicator's expectation determines the expectation of its complement:

```
  EX≈PrX⇒EXᶜ≈PrXᶜ : E X ≈ᴹ PR I *ₗ 1ᴹ → E Xᶜ ≈ᴹ PR (C I) *ₗ 1ᴹ
  -- proof elided
```

### 3.4.1 Satisfiable Indicators

The `Satisfiable` condition identifies nontrivial indicators whose event and complement both have strictly positive probability. This condition ensures that conditioning on the event is well defined.

```
Satisfiable : Indicator S → Type
Satisfiable I = 0# < PR I × 0# < PR (C I)
```

The `Satisfiable` condition will be used in Section 3.5.0.1 to implement conditioning on an indicator event. Conditional random variables are defined more generally, but with indicator function a concrete implementation can be constructed.

## 3.5 Conditional Random Variables

A conditional random variable is a relation between two random variables which formalizes how the first random variable should be updated "given" the second random variable.

```
record E[_|_] (Y X : RV) : Type₁ where
```

Specifically, the conditional random variable relation states the existence of another random variable, denoted `E[Y|X]`, which must satisfy two conditions.

```
    E[Y|X] : RV
```

The `X-meas` condition requires that the conditional random variable depends only on the value of `X`. Outcomes yielding the same value of `X` must yield the same conditional random variable.

```
    X-meas : ∀ (ω₁ ω₂ : Ω) → to X ω₁ ≈ᴹ to X ω₂ → to E[Y|X] ω₁ ≈ᴹ to E[Y|X] ω₂
```

The orthogonality condition requires that the difference `Y - E[Y|X]` is orthogonal to any function of `X`, making the conditional random variable the best predictor of `Y` using only information about `X`.

```
    orthogonal : ∀ (g : Mₛ → Mₛ) → ⟨ Y -′ E[Y|X] , g ∘→ X ⟩ ≈ᴹ 0ᴹ
```

These conditions uniquely characterize conditional random variables. In Section 2.4, it was shown that `E[Y|X]` implies the law of total expectation.

#### 3.5.0.1 Conditioning on Indicators

Constructing an `E[_|_]` type is challenging in general, but below broadly applicable theorems for conditioning on `Satisfiable Indicator` random variables are proven and demonstrated in various examples in Section 5. First, the `condI` function defines the random variable which implements conditioning:

```
condI : RV → ∃ Satisfiable → RV
to (condI Y (I , 0<PrI , 0<PrCI)) ω with isIndicator I ω
    -- When I ω ≈ 1ᴹ, scale E (I *′ Y) by 1/ PrI
... | inj₁ _ = 1/ (ℛ.#-sym (ℛ.<⇒# 0<PrI)) *ₗ ⟨ I I , Y ⟩
    -- When I ω ≈ 0ᴹ, scale E (C I *′ Y) by 1/ PrCI
... | inj₂ _ = 1/ (ℛ.#-sym (ℛ.<⇒# 0<PrCI)) *ₗ ⟨ I (C I) , Y ⟩
--- proof of congruence elided
```

While `condI` function constructs the `E[Y|X]` random variable, then it remains to show that `condI` satisfies the `X-meas` and `orthogonal` properties. The `condIMeas` theorem proves `X-meas` for `condI` by pattern-matching on the `isIndicator` property of indicator functions (they yield either a 1 or 0):

```
module _ (Y : RV) (X@(I , _) : ∃ Satisfiable) where

  condIMeas : ∀ (ω₁ ω₂ : Ω)
      → to (toRV I) ω₁ ≈ᴹ to (toRV I) ω₂
      → to (condI Y X) ω₁ ≈ᴹ to (condI Y X) ω₂
  condIMeas ω₁ ω₂ Xω₁≈Xω₂ with isIndicator I ω₁ | isIndicator I ω₂
  ... | inj₁ _     | inj₁ _     = Eq.refl
  ... | inj₂ _     | inj₂ _     = Eq.refl
  ... | inj₁ Iω₁≈1 | inj₂ Iω₂≈0 = ⊥-elim
```

```
    (<ᴹ-irrefl
        (Eq.trans (Eq.trans (Eq.sym Iω₂≈0) (Eq.sym Xω₁≈Xω₂)) Iω₁≈1)
        <ᴹ-nonTrivial)
  ... | inj₂ Iω₁≈0 | inj₁ Iω₂≈1 = ⊥-elim
      (<ᴹ-irrefl
        (Eq.trans (Eq.trans (Eq.sym Iω₁≈0) Xω₁≈Xω₂) Iω₂≈1)
        <ᴹ-nonTrivial)
```

The `condOrth` theorem stated below proves `orthogonal` for `condI`, under the condition that for an indicator `I`, `E (toRV I) ≈ᴹ PR I *ι 1ᴹ`. This condition is trivial when the carrier type of the `MAlgebra` is scalars, as the trace operation is the identity function in that case. The proof of `condOrth` involves several helper lemmas and multiple algebraic manipulations, so it is elided from the manuscript.

```
module _ (Y : RV) (Xi@(I , 0<PrI , 0<PrCI) : ∃ Satisfiable) (g : Mₛ → Mₛ) where

  condOrth : E (toRV I) ≈ᴹ PR I *ι 1ᴹ → ⟨ Y -′ condI Y Xi , g ∘→ toRV I ⟩ ≈ᴹ 0ᴹ
```

With the `condI` function, and `condIMeas` and `condOrth` theorems, `mkEY|I` functions constructs a conditional random variable from `Y` “given” `X`.

```
module _ (Y : RV) (Xi@(I , 0<PrI , 0<PrCI) : ∃ Satisfiable) where

  private
    PrX   = PR I
    X     = toRV I
    EY|X = condI Y Xi

  mkEY|I : E X ≈ᴹ PrX *ι 1ᴹ → E[ Y | X ]
  mkEY|I EX≈PrX = mk EY|X (condIMeas Y Xi) λ g → condOrth Y Xi g EX≈PrX
```

### 3.6 Denotation, Equivalence, and Ordering

Before defining the primitive operations for building $\mathcal{E}$ spaces, a basic notion of equivalence and ordering of $\mathcal{E}$ over the same sample space is defined. The `≈ᴱ` relation defined below is used in Section 4 to specify the laws of the primitive operations. Both `≈ᴱ` and `≤ᴱ` can be shown (elided) to be equivalence and partial order relations, respectively.

The denotation function defines the semantics of an expectation space by extracting the expectation operator from an instance of the expectation type:

```
⟦_⟧ : ℰ S → (S → Mₛ) → M
⟦_⟧ = ℰ.E
```

Two expectation spaces are considered equivalent if their meanings are equivalent; that is, if they assign the same expected value to all random variables:

```
_≈ᴱ_ : ℰ S → ℰ S → Type
_≈ᴱ_ = (λ E₁ E₂ → ∀ {X} → E₁ X ≈ᴹ E₂ X) on ⟦_⟧
```

Similarly, two expectation spaces are ordered by the ordering of their meanings:

```
_≤ᴱ_ : ℰ S → ℰ S → Type
_≤ᴱ_ = (λ E₁ E₂ → ∀ {X} → E₁ X ≤ᴹ E₂ X) on ⟦_⟧
```

Since ≈ᴱ is an equivalence relation, $\mathcal{E}$ spaces form a `Setoid`, denoted $\mathcal{E}_s$:

```
ℰₛ : SampleSpace → Setoid (lsuc 0ℓ) 0ℓ
ℰₛ S = record { Carrier = ℰ S ; isEquivalence = ≈ᴱ-isEquivalence }
```

In this paper, in order to keep the explication simpler, `SampleSpace`s are `Setoid`s restricted to universe level `0ℓ`. In principle, however, the $\mathcal{E}$ could take `Setoid`s at any level, including `ℰₛ S`, for same space `S`.

## 3.7 Markov Kernel

The final definition needed before introducing the core constructs for building expectations in Section 4 is a Markov Kernel:

```
_→ℰ_ : SampleSpace → SampleSpace → Type₁
S₁ →ℰ S₂ = S₁ → ℰₛ S₂
```

This mathematical object, which assigns to each outcome in one sample space an expectation over another (possibly different) sample space, appears under various names throughout the statistical and categorical probability literature. Throughout the literature, this type is referred to by various names, including:

- Markov kernel in the theory of Markov chains and stochastic processes
- Transition kernel in the study of stochastic dynamics
- Statistical model or likelihood in statistical inference
- Conditional distribution in probability theory
- Kleisli arrow in the categorical approach to probability

This definition captures the essential structure underlying many statistical concepts. For example, Markov kernels enable modeling of:

- Transition probabilities in Markov chains, where the distribution of the next state depends on the current state;
- Conditional distributions in hierarchical models, where parameters at one level determine distributions at subsequent levels;
- Likelihood functions in statistical inference, where the distribution of observations depends on unknown parameters;
- Random functions in functional data analysis, where each input value is associated with a probability distribution over outputs;
- Measurement models where the distribution of observed data depends on underlying latent states.

In all these cases, the underlying object is the same: a function assigning a distribution to each input. What varies is how this structure is interpreted and used.

The power of Markov kernels lies in their compositional nature. Complex probabilistic models can be constructed by composing simpler kernels. This compositional approach, explored in detail in Section 4.3.1 and Section 5, provides a principled foundation for building sophisticated statistical models while maintaining formal correctness guarantees. The bind operation >>= introduced in Section 4.1.4 demonstrates the role of Markov kernels in statistical modeling. When an expectation space `𝔼 : ℰ S₁` is composed with a kernel `f : S₁ →ℰ S₂` via the bind operation `𝔼 >>= f`, the result is a new expectation space over `S₂`. This operation enables the construction of sequential processes where later stages depend on the outcomes of earlier stages.

This section formalized the concept of expectation as a primitive notion, defining the necessary properties that characterize statistical expectations. The $\mathcal{E}$ type provides a foundation for statistical reasoning and sets the groundwork for the theorems and examples explored in subsequent sections. The preceding section introduced statistical concepts such as probability and conditional random variables, expressed in terms of a single expectation space. The next step is to equip $\mathcal{E}$ with operations that construct new expectation spaces from existing ones. These primitive operations serve as the building blocks for assembling statistical models.

# 4 Building Expectations: Building Blocks of Distributions

Statistical modeling involves combining simple probabilistic building blocks into complex distributions. Just as programming languages provide primitive operations that can be combined to create sophisticated programs, statistical theory requires primitive operations for distributions that can be composed to build realistic statistical models.

This section establishes five fundamental operations on $\mathcal{E}$ spaces that serve as the foundation for statistical distributions: creating deterministic outcomes (`pure`), transforming sample spaces (`map`), lifting function application to the $\mathcal{E}$ level (`⟨*⟩`), creating dependent sequences (>>=), and forming mixture distributions (`mix`). These operations provide a foundation for constructing statistical models. The operations `pure`, `map`, `⟨*⟩`, and >>= derive from category theory, specifically the theory of functors, applicative functors, and monads (Mac Lane 1998; McBride and Paterson 2008), though this connection is not dwelled upon here.

The following operations show only the definition for the expectation operator `E`. The formal proofs that these operations satisfy required mathematical properties (such as linearity, positivity, and normalization) are elided from the presentation but included in the complete source code.

## 4.1 Primitive operations on $\mathcal{E}$

### 4.1.1 Pure deterministic expectation

The `pure` function creates an expectation operator that evaluates each random variable at the given point `ω`. The `pure` function corresponds to the Dirac measure, which places all probability mass on a single outcome.

```
pure : Ω S → ℰ S
pure ω = mk (λ X → to X ω)
```

### 4.1.2 Transforming Sample Spaces

The `map` operation takes a function `S₁ → S₂` and a distribution over sample space `S₁` to induce a distribution on the transformed space. Many of the derived operations in Section 4.2 and the distributions constructed in Section 5 are simple applications of `map`.

```
map : S₁ → S₂ → ℰ S₁ → ℰ S₂
map f 𝔼 = mk (λ X → E (X ∘→ f))
```

An infix version of `map`, denoted `_⟨$⟩_`, is also provided, so `map f 𝔼` is syntactically equivalent to `𝔼 ⟨$⟩ f`.

### 4.1.3 Expectation of Function Application

The `_⟨*⟩_` operator combines a distribution over a sample space of functions (`S₁ →ₛ S₂`) with a distribution over `S₁` to produce a distribution over `S₂`. This enables modeling scenarios where both the transformation being applied and the values being transformed have a distribution. The `_⟨*⟩_` is pivotal to defining `map₂` in Section 4.2, which lifts binary sample space transformations.

```
_⟨*⟩_ : 𝓔 (S₁ →ₛ S₂) → 𝓔 S₁ → 𝓔 S₂
𝔽 ⟨*⟩ 𝔼 = mk
  (λ X → 𝔽.E (mk→ (λ f → 𝔼.E (X ∘→ f)) λ x → 𝔼.E-cong λ {ω} → cong X (x {ω})))
```

### 4.1.4 Sequentially Composing Expectations

The bind operator _>>=_ operation provides the primitive mechanism for constructing any process of expectations where later stages depend on earlier outcomes. This models scenarios where the expectation of subsequent random variables depends on the realized values of previous ones.

```
_>>=_ : 𝓔 S₁ → (S₁ →𝓔 S₂) → 𝓔 S₂
𝔼 >>= f = mk
  (λ X → 𝔼.E (mk→ (λ ω → 𝓔.E (to f ω) X) (λ ω≈ω′ → cong f ω≈ω′)))
```

### 4.1.5 Mixing Expectations

While the previous operations are tools for updating or modifying expectations, the `mix` function defines the fundamental operation of combining expectations. Every probabilistic model that involves uncertainty about which of several scenarios will occur relies the `mix` operation, from the simplest coin flip to the most complex hierarchical model.

```
mix : ∀ {S : Fin⁺ n → SampleSpace}
  → ((i : Fin⁺ n) → 𝓔 (S i))
  → Simplex n
  → 𝓔 (∐[ n ] S)
mix {n = 1 #}              f _ = f zero
mix {n = n +2 #} {S = S} f s = mk
  (λ X → ∑ λ i → vs i *ₗ 𝓔.E (f i) (X ∘→ injᵢ {n = n +2 #} {S = S} i))
```

## 4.2 Derived Operations on $\mathcal{E}$

The primitive operations established in the previous section provide the foundation for constructing more sophisticated expectation operations. This section develops derived operations that capture common statistical concepts such as: pushing distributions through transformations, combining independent distributions, extracting marginal distributions, and creating complex mixture models. These derived operations demonstrate the compositionality of the expectation framework. Complex statistical constructions emerge naturally from systematic combination of simpler primitives.

### 4.2.1 Basic Operations

The degenerate distribution `𝓔𝟙` is an expectation operator over the unit type, which has only one possible outcome. This serves as a building block for more complex distributions.

```
𝔼𝟙 : 𝓔 𝟙
𝔼𝟙 = pure tt
```

The `map₂` function lifts binary sample space transformations to transformations of expectations over those sample spaces, enabling the combination of random variables with different sample spaces. It underlies a wide range of statistical operations, from basic arithmetic on random variables to complex functional relationships between distributions.

```
map₂ : S₁ → S₂ → S₃ → ℰ S₁ → ℰ S₂ → ℰ S₃
map₂ _·_ 𝔼 = map _·_ 𝔼 ⟨*⟩_
```

### 4.2.2 Joint Distribution Operations

The product operation _×_ creates joint distributions from independent component distributions. This construction enables modeling scenarios where multiple random processes operate independently. The resulting distribution satisfies the independence property: the joint probability factorizes as the product of marginal probabilities.

```
_×_ : ℰ S₁ → ℰ S₂ → ℰ (S₁ ×ₛ S₂)
_×_ = map₂ pair
```

The $\pi_1$ and $\pi_2$ operations extract the marginal distribution of the first component (resp. second) from a joint distribution.

```
π₁ : ℰ (S₁ ×ₛ S₂) → ℰ S₁
π₁ = map proj₁

π₂ : ℰ (S₁ ×ₛ S₂) → ℰ S₂
π₂ = map proj₂
```

The ℰ∏ function constructs a joint distribution over a product of different sample spaces. This construction ensures that the distributions are independent, meaning the joint probability factorizes as the product of the individual probabilities.

```
ℰ∏ₙ : (n : ℕ⁺) → {Sᵛ : Fin⁺ n → SampleSpace}
  → ((i : Fin⁺ n) → ℰ (Sᵛ i)) → ℰ (∏[ n ] Sᵛ)
ℰ∏ₙ (1 #) f = f zero
ℰ∏ₙ (n +2 #) f = f zero × ℰ∏ₙ (n +1 #) (f ∘ suc)
```

The ∏ function specializes ℰ∏ to the case where all component sample spaces are identical:

```
∏ₙ : (n : ℕ⁺) → ℰ S → ℰ (S ^ n)
∏ₙ n = ℰ∏ₙ n ∘ const
```

The $\pi_i$ function extracts the marginal distribution of the i-th component from a joint distribution:

```
πᵢ : ℰ (∏[ n ] Sᵛ) → (i : Fin⁺ n) → ℰ (Sᵛ i)
πᵢ 𝔼 i = 𝔼 ⟨$⟩ (projᵢ i)
```

### 4.2.3 Algebraic Operations: Lifting Structure to Distributions

The expectation framework enables systematic lifting of algebraic structures from sample spaces to distributions over those spaces. Algebraic structure on a sample space induces a corresponding structure on distributions over that space. The convolution operation _⊙_ below exemplifies this lifting principle. Given any monoid structure on a sample space, convolution lifts the monoid operation to distributions:

```
module ConvolutionMonoid (M : Monoid 0ℓ 0ℓ) where
```

```
open module M = Monoid M

Monₛ : SampleSpace
Monₛ = M.setoid

_⊙_ : ℰ Monₛ → ℰ Monₛ → ℰ Monₛ
_⊙_ = map (mk→ (×.uncurry _•_) (×.uncurry M.•-cong)) ∘₂ _×_

ℰε : ℰ Monₛ
ℰε = pure ε
```

This construction creates a distribution over the monoid operation applied to independent pairs. The resulting distributions themselves form a monoid (proof elided), with convolution as the operation and the unit distribution `pure ε` as the identity.

This pattern generalizes beyond monoids. Algebraic structures such as semigroups, groups, rings, or lattices can be lifted to the corresponding structure on distributions. The primitive operations provide the necessary compositional tools to transfer algebraic properties from sample spaces to distributions over those spaces. For example, since every `MAlgebra` contains a monoid, distributions over the measurement space automatically inherit convolution:

```
_⊕_ : ℰ Mₛ → ℰ Mₛ → ℰ Mₛ
_⊕_ = ConvolutionMonoid._⊙_ +ᴹ-monoid
```

### 4.2.4 Mixture Operations

The mixture operations provide systematic ways to combine multiple expectations. These operations generalize the basic `mix` function to handle common statistical scenarios involving discrete choices, uniform combinations, and empirical distributions.

#### 4.2.4.1 Binary Mixtures

The `_+[_]_` operation creates a mixture over different sample spaces, combining different types of outcomes.

```
_+[_]_ : ℰ S₁ → 𝑃 → ℰ S₂ → ℰ (S₁ +ₛ S₂)
_+[_]_ {S₁} {S₂} 𝔼₁ p 𝔼₂ = mix
  {S = λ { 0F → S₁ ; 1F → S₂ }}
  (λ { 0F → 𝔼₁ ; 1F → 𝔼₂ })
  (𝑃⇒Simplex p)
```

The `choose` function specializes `_+[_]_` to the case where both component distributions are over the same sample space, combining of homogeneous alternatives.

```
choose : 𝑃 → ℰ S → ℰ S → ℰ S
choose p = map reduce ∘₂ _+[ p ]_
```

#### 4.2.4.2 N-ary Mixtures

The n-ary constructor `mixₙ` packages a single expectation `𝔼 : ℰ S` with a weight vector `s : Simplex n` to produce a distribution on the tagged `copower⁺ n S` (i.e., n labeled copies of S).

Concretely, `mixₙ` draws an index `i` from `Fin⁺ n`, a non-empty finite set with probability $s_i$ and then samples according to 𝔼, tagging the outcome with `i`.

```
mixₙ :  ℰ S → Simplex n → ℰ (copower⁺ n S)
mixₙ 𝔼 = mix (const 𝔼)
```

The `uniformComponentMixture` function creates a distribution over a collection of heterogeneous tuples. Given a function that produces `m` different tuples, each containing components from potentially different sample spaces, it returns a distribution that assigns equal probability to each tuple. This operation is particularly useful for creating empirical distributions from multivariate observations where different dimensions may have different types, enabling the construction of joint distributions over diverse parameter spaces.

```
uniformComponentMixture : ∀ {S : Fin⁺ n → SampleSpace} {m}
  → (Fin⁺ m → ℰ (∏[ n ] S))
  → ℰ (∏[ n ] S)
uniformComponentMixture {n} {S} {m = 1 #} f = f zero
uniformComponentMixture {n} {S} {m = m +2 #} f =
  choose 1/[ₙ m +2 ]
  (f zero)
  (uniformComponentMixture {n} {S} (f ∘ suc))
```

The `uniformTupleMixture` function, on the other hand, deals directly with values rather than distributions. In practical terms, `uniformComponentMixture` mixes arbitrary distributions, while `uniformTupleMixture` creates a discrete distribution over a specific collection of tuples.

```
uniformTupleMixture : ∀ {S : Fin⁺ n → SampleSpace} {m : ℕ⁺}
  → (Fin⁺ m → (∀ i → Ω (S i)))
  → ℰ (∏[ n ] S)
uniformTupleMixture {n} {S} {m} f =
  uniformComponentMixture {n} {S} {m} (pure ∘ toTuple ∘ f)
```

The `uniformMixture` function creates a uniform distribution over a collection of values from the same sample space. It takes a vector of length `n` containing elements from sample space `S` and returns a distribution that assigns equal probability (1/n) to each element. This implementation leverages the more general `uniformTupleMixture` function by treating the homogeneous case as a special case of the heterogeneous one.

```
uniformMixture : (Fin⁺ n → (Ω S)) → ℰ S
uniformMixture {n} {S} f =
  uniformTupleMixture {n = 1 #} {S = const S} {m = n} λ i _ → f i
```

## 4.3 Operations on Kernels

The preceding sections defined operations that act directly on expectation spaces. Many statistical models, however, require transformations that depend on the specific input value. Markov kernels (Section 3.7) provide this capability by mapping each element of one sample space to an expectation space over another. The following section develops operations on kernels for composing such transformations into sequential, hierarchical, and conditional models.

The identity kernel $\mathtt{id}_k$ represents the trivial probabilistic transformation that maps each outcome to a degenerate distribution concentrated on that same outcome. This serves as the identity element for kernel composition:

#### 4.3.1 Basic Kernel Operations

```
idₖ : S →ℰ S
```

The composition operation $_\circ_k_$ enables sequential application of probabilistic transformations. When kernel $\mathtt{g}$ maps sample space $S_1$ to distributions over $S_2$, and kernel $\mathtt{f}$ maps $S_2$ to distributions over $S_3$, their composition $\mathtt{f} \circ_k \mathtt{g}$ produces a kernel from $S_1$ to $S_3$. This operation can be used for modeling multi-stage probabilistic processes:

```
_∘ₖ_ : S₂ →ℰ S₃ → S₁ →ℰ S₂ → S₁ →ℰ S₃
```

#### 4.3.2 Product Operations

The product operations enable modeling of scenarios involving jointly distributed components. The diagonal operation $_\vartriangle_k_$ takes two kernels from the same source space and produces a kernel that generates pairs by applying both kernels to the same input:

```
_▵ₖ_ : S₁ →ℰ S₂ → S₁ →ℰ S₃ → S₁ →ℰ (S₂ ×ₛ S₃)
```

The tensor product $_\otimes_k_$ operates on pairs of inputs, applying each kernel to its respective component. This enables modeling of scenarios where different components undergo independent probabilistic transformations:

```
_⊗ₖ_ : S₁ →ℰ S₃ → S₂ →ℰ S₄ → (S₁ ×ₛ S₂) →ℰ (S₃ ×ₛ S₄)
```

Projection kernels extract components from joint sample spaces. The operations $\mathtt{proj}_{1k}$ and $\mathtt{proj}_{2k}$ extract the first and second components from pairs, respectively:

```
proj₁ₖ : (S₁ ×ₛ S₂) →ℰ S₁
proj₂ₖ : (S₁ ×ₛ S₂) →ℰ S₂
```

These operations are used for marginalizing joint distributions and extracting individual components from multivariate models.

#### 4.3.3 Sum Operations

The case analysis operation $_\triangledown_k_$ applies the appropriate kernel based on which alternative is present:

```
_▿ₖ_ : S₁ →ℰ S₃ → S₂ →ℰ S₃ → (S₁ +ₛ S₂) →ℰ S₃
```

The sum operation $_\oplus_k_$ transforms both alternatives independently:

```
_⊕ₖ_ : S₁ →ℰ S₂ → S₃ →ℰ S₄ → (S₁ +ₛ S₃) →ℰ (S₂ +ₛ S₄)
```

The injection kernels $\mathtt{inj}_{1k}$ and $\mathtt{inj}_{2k}$ embed sample spaces into sum types:

```
inj₁ₖ : S₁ →ℰ (S₁ +ₛ S₂)
inj₂ₖ : S₂ →ℰ (S₁ +ₛ S₂)
```

#### 4.3.4 Other Operations

The conditional operation $\mathtt{if}_k\mathtt{_Then_Else_}$ enables probabilistic branching based on binary tests:

```
ifₖ_then_else_ : S₁ →ℰ 2 → S₁ →ℰ S₂ → S₁ →ℰ S₂ → S₁ →ℰ S₂
```

This operation first applies a test kernel that produces a binary outcome, then applies either the "then" or "else" kernel based on the result. This construction can be used for modeling decision processes, survival analysis with competing risks, and any scenario where the probabilistic behavior depends on observable conditions.

The fixed point approximation `fix-approxₖ` enables modeling of iterative or recursive probabilistic processes:

```
fix-approxₖ : ℕ → (S₁ →ℰ S₂) → (S₁ →ℰ S₂ → S₁ →ℰ S₂) → (S₁ →ℰ S₂)
```

This operation takes a fuel parameter (limiting iterations), an initial kernel, and a transformation of kernels, producing an approximation to the fixed point. This operation is used in modeling processes that continue until some termination condition is met, such as sequential sampling schemes, iterative algorithms with random components, or epidemic models with feedback loops.

# 5 Expecting Examples

With a vocabulary of operations on both expectation spaces and kernels in place, the framework is now ready to generate complete statistical models. The following section constructs familiar named distributions and more complex examples by systematically composing operations defined in Section 4, demonstrating how the algebraic laws of $\mathcal{E}$ and the compositional structure of kernels guarantee correctness by construction.

In traditional statistics, probability distributions are often introduced via named formulas: pmfs for discrete cases, pdfs for continuous ones. In the framework presented here, such functions emerge as consequences of composition. A distribution is simply an expectation space ($\mathcal{E}$ `S`), built from smaller elements: point masses, random choices, and sequencing via kernels. For example:

- A Bernoulli distribution is a mixture of two outcomes chosen by a coin flip.
- A Binomial arises by iterating that choice.
- More complex models (e.g., Poisson, Negative Binomial) are constructed similarly.

The familiar mass and density functions are not primitive. Instead, they are theorems that follow from the algebraic structure, which ensures normalization, consistency, and composability by construction. This compositional view shifts focus from closed-form formulas to the generative structure behind distributions, enabling systematic construction, transformation, and analysis of probabilistic models.

## 5.1 Choice

Statistical modeling frequently involves scenarios where outcomes are selected from a finite set of alternatives. From the basic coin flip to categorical variables in survey data, the mathematics of choice underpins many common models. This section develops such choice-based models using the expectation framework, showing how simple choice mechanisms give rise to familiar discrete distributions. The key insight is that selecting among alternatives with specified probabilities can be understood as a mixture over outcome spaces. Rather than treating each named distribution (e.g., Bernoulli, Categorical) as a separate object, this framework exposes the shared algebraic structure beneath them.

### 5.1.1 $k$ partitions

One important class of choice models is choosing between $k : \mathbb{N}^{+}$ distinct options. A `Simplex k` value specifies a convex weighting over these $k$ options, and can be used to construct an expectation space on the corresponding finite set. The `kparts` function below takes such a simplex and produces an expectation over the finite index type `Fin⁺ₛ k`:

```
kparts : Simplex k → ℰ (Fin⁺ₛ k)
kparts s = mixₙ 𝔼𝟙 s ⟨$⟩ ∐𝟙→Fin
```

This construction uses the `mixₙ` operation (introduced in Section 4.2.4) to combine $k$ copies of the unit expectation `𝔼𝟙`, each representing a single outcome, weighted according to the components of the simplex `s`. The function `∐𝟙→Fin` reshapes the result from a tagged sum to the canonical index type `Fin⁺ₛ k`. The resulting expectation space reflects a randomized selection among $k$ possibilities, with weights specified by `s`.

### 5.1.2 Discrete Uniform

The discrete uniform distribution represents the case where all alternatives are equally likely. This distribution arises naturally as a special case of `kparts`:

```
discreteUniform : (n : ℕ⁺) → ℰ (Fin⁺ₛ n)
discreteUniform = kparts ∘ uniformSimplex
```

```
𝔼 : ℰ (Fin⁺ₛ (4 #))
𝔼 = discreteUniform (4 #)
```

For a 4-element discrete uniform distribution, one can verify that each outcome receives probability exactly $\frac{1}{4}$:

```
_ : PR (δF 0F) ≟ + 1 / 4 ; _ = ✓
_ : PR (δF 1F) ≟ + 1 / 4 ; _ = ✓
_ : PR (δF 2F) ≟ + 1 / 4 ; _ = ✓
_ : PR (δF 3F) ≟ + 1 / 4 ; _ = ✓
```

### 5.1.3 Discrete Uniform over ℤ

Choice distributions naturally generalize beyond finite index sets to other discrete spaces. The `discreteUniformℤ` function demonstrates transforming the sample space for `discreteUniform` to the support of the discrete uniform typically found in textbooks:

```
discreteUniformℤ : ℤ → ℕ⁺ → ℰ ℤₛ
discreteUniformℤ a n = discreteUniform n ⟨$⟩ mk→ᶠ (λ i → a ℤ.+ + (toℕ i))
```

This construction first creates a discrete uniform distribution over `n` finite alternatives, then applies the map operation (`⟨$⟩`) to transform each finite index into an integer. The mapping function `λ i → a ℤ.+ + (toℕ i)` translates index `i` into the integer `a + i`, creating a uniform distribution over the consecutive integers $\{a, a + 1, ..., a + n - 1\}$.

Below, the distribution from 0…5 is constructed, and its expectation and variance computed exactly:

```
𝔼 = discreteUniformℤ (0ℤ) (6 #)
```

```
-- Mean: (5 + 0) / 2 = 2.5
_ : E ℤ⇒ℚᵘ ≟ + 5 / 2 ; _ = ✓

-- Variance: ((5 + 0 + 1)² - 1) / 12 = 35/12
_ : Var ℤ⇒ℚᵘ ≟ + 35 / 12 ; _ = ✓
```

The ℤ⇒ℚᵘ random variable maps each integer in the sample space to its corresponding rational value. It lets expectations over ℤₛ be computed in ℚᵘₛ, using the canonical embedding.

The same distribution can be conditioned on a predicate such as $X \leq 2$, using an indicator:

```
I : Indicator ℤₛ
I = 𝟙≤ℤ (+ 2)
```

Once shown satisfiable:

```
IisSatisfiable : Satisfiable I
```

the conditional random variable can be constructed:

```
ex : E[ ℤ⇒ℚᵘ | toRV I ]
ex = mkEY|I ℤ⇒ℚᵘ (I , IisSatisfiable) ℚᵘ.≃-refl
```

This defines a function `E[Y|X] : ℤₛ → ℚᵘₛ`, which returns the conditional mean depending on whether the input lies within the indicator set:

```
_ : to E[Y|X] (+ 2)          ≟ + 1 / 1 ; _ = ✓
_ : to E[Y|X] (ℤ.- (+ 10))   ≟ + 1 / 1 ; _ = ✓
_ : to E[Y|X] (+ 10)         ≟ + 4 / 1 ; _ = ✓
```

These computations verify that values in the range $X \leq 2$ yield the correct conditional mean of 1, while values outside yield the complement's mean of 4.

### 5.1.4 Bernoulli

The Bernoulli distribution models the simplest nontrivial choice: a binary outcome selected according to a proportion $p$.

```
bernoulli : P → ℰ 2
bernoulli p = 𝔼1 +[ p ] 𝔼1
```

This construction uses the binary mixture operation `_+[_]_` to combine two unit expectations 𝔼1 with proportion `p` for the first outcome and proportion $1 - p$ for the second. The type 2 represents the two-element Boolean space, capturing binary outcomes. A kernel version lifts `bernoulli` to a Markov kernel from proportions to an expectation space over 2:

```
bernoulliKernel : Pₛ →ℰ 2
```

#### 5.1.4.1 Mixtures of Bernoulli Distributions

The framework extends to mixtures of Bernoulli distributions, where the success probability is itself treated as a random variable. Consider a mixture over three possible values of `p`, each with specified weight:

```
𝔼P : ℰ Pₛ
𝔼P = uniformMixture {n = 3 #}
  λ { 0F → 1/[ₙ 2 ] -- 1/2
    ; 1F → 1/[ₙ 3 ] -- 1/3
    ; 2F → 1/[ₙ 4 ] -- 1/4
    }
```

```
𝔼 : ℰ 2
𝔼 = 𝔼P >>= bernoulliKernel
```

This creates a distribution over 2 by drawing a value of p from 𝔼P and then drawing a Bernoulli outcome with that p. The marginal success probability is computed as:

```
-- Marginal probability of success for the mixture:
_ : ℰ.E 𝔼 (1ℚᵘ ⟨▿⟩ 0ℚᵘ)  ≟ + 13 / 36 ; _ = ✓
```

This matches the expected value under the explicit weighted sum:

$$P(X = 1) = \sum P(X = 1 \mid p) \times P(p) = \frac{1}{2} \times \frac{1}{3} + \frac{1}{3} \times \frac{1}{3} + \frac{1}{4} \times \frac{1}{3} = \frac{1}{6} + \frac{1}{9} + \frac{1}{12} = \frac{13}{36}$$

#### 5.1.4.2 Posterior Inference with Joint Distributions

To compute posterior expectations of the proportions in 𝔼P given a Bernoulli outcome, define the joint distribution over both $P_s$ and 2 by updating 𝔼P with bernoulliKernel:

```
𝔼′ : ℰ (𝑃ₛ ×ₛ 2)
𝔼′ = 𝔼P >>= (pureKernel ▵ₖ bernoulliKernel)
```

The random variable P extracts the $P$ component from the product space:

```
P : (𝑃ₛ ×ₛ 2) → ℚᵘₛ
P = mk→ 𝑃.p (λ x → x) ∘→ proj₁
```

Define an indicator random variable that maps the second component of the pair to 1 if it is the left injection (interpreted as "true") of 2 and to 0 otherwise. This construction embeds the Boolean structure of 2 as an indicator in ℚ:

```
X : Indicator (𝑃ₛ ×ₛ 2)
```

Construct the conditional random variable of P given the event encoded by the indicator X:

```
ex : E[ P | toRV X ]
ex = mkEY|I P (X , XSatisfiable) ℚᵘ.≃-refl
```

The resulting conditional random variable has type (𝑃ₛ ×ₛ 2) → ℚᵘₛ. Since the indicator depends only on the second component, the conditional random variable is constant in the first argument. Therefore, it can be reinterpreted as a function from 2 to ℚᵘₛ by fixing any value of $P$:

```
E[P|X] : 2 → ℚᵘₛ
to E[P|X] x = to E[Y|X] (1/[ℕ 1 ] , x)
cong E[P|X] x = cong E[Y|X] (ℚᵘ.≃-refl  , x)
```

To confirm the computational content of the E[P|Y] random variable, the following values are the posterior means of p after observing success (⊎.inj₁) or failure (⊎.inj₂).

```
_ : to E[P|X] (⊎.inj₁ tt) ≟ + 61 / 156 ; _ = ✓
_ : to E[P|X] (⊎.inj₂ tt) ≟ + 95 / 276 ; _ = ✓
```

### 5.1.5 Binary Urn

A simple urn distribution models drawing a single item from an urn containing a known number of items of two types. Given `m` items of one type and `n` of the other, the resulting expectation space represents a Bernoulli trial with success probability `m / (m + n)`.

```
binaryUrn : ℕ → ℕ → ℰ 2
binaryUrn = bernoulli ∘₂ m/[m+n]
```

This construction uses the `bernoulli` function to mix two outcomes with the appropriate frequency ratio, applying the `m/[m+n]` operation to convert counts into a valid proportion.

## 5.2 Repetition

Many statistical models arise from repeating a process: flipping a coin multiple times, drawing from an urn, or sampling until a condition is met. This section demonstrates how such models can be constructed compositionally.

### 5.2.1 Independent trials

The simplest form of repetition occurs when experiments are conducted independently, with each trial assumed to follow the same distribution regardless of previous outcomes.

#### 5.2.1.1 k-partitions trials

This construction models `n` independent draws from a `k`-way choice, where each outcome is selected according to weights in a `Simplex`.

```
kpartsTrials : (n : ℕ⁺) → Simplex k → ℰ ((Fin⁺ₛ k) ^ n)
kpartsTrials n = ∏ₙ n ∘ kparts
```

Independence arises from the repeated product `∏ₙ`, which constructs a distribution over sequences by composing identical trials without dependence. Each trial draws from `Fin⁺ₛ k` according to the weights in the simplex.

#### 5.2.1.2 Bernoulli trials

The special case of binary choice leads to the classical Bernoulli trials:

```
bernoulliTrials : (n : ℕ⁺) → P → ℰ (2 ^ n)
bernoulliTrials n = ∏ₙ n ∘ bernoulli
```

This construction creates a distribution over binary sequences of length `n`, where each position has success probability `p` and failure probability `1 - p`, independently of all other positions. The type `2 ^ n` represents the space of all possible binary sequences of length `n`, capturing the complete sample space for `n` independent binary trials.

### 5.2.2 Dependent Trials: State-Dependent Processes

Many scenarios involve dependence between trials, where the outcome of current experiments influences the expectations of future outcomes.

#### 5.2.2.1 Urn trials

The urn model is a classic example of dependent trials: sampling without replacement creates statistical dependence between draws.

```
urnTrials : (s f : ℕ) → (n : ℕ⁺) → ℰ (2 ^ n)
urnTrials s f (1 #) = binaryUrn s f
```

```
urnTrials 0 0 (n +2 #) = ∏ₙ (n +2 #) (binaryUrn 0 0)
urnTrials (s +1) 0 (n +2 #) = urnTrials s 0 (n +1 #) ⟨$⟩ inj₁⁺ {n}
urnTrials 0 (f +1) (n +2 #) = urnTrials 0 f (n +1 #) ⟨$⟩ inj₂⁺ {n}
urnTrials (s +1) (f +1) (n +2 #) =
  binaryUrn (s +1) (f +1) >>=
    K (urnTrials s (f +1) (n +1 #) ⟨$⟩ inj₁⁺ {n}) ▿
    K (urnTrials (s +1) f (n +1 #) ⟨$⟩ inj₂⁺ {n})
```

This models an urn containing `s` success balls and `f` failure balls, sampled without replacement:

- Base case (`n = 1`): Draw once from the current urn.
- Empty urn cases: When only one type of ball remains, all remaining draws have deterministic outcomes.
- General case: Draw once, then update the urn and recurse.

The key step is recursing bind (>>=), which conditions future draws on the current outcome. The ▿ choice operator routes control flow based on whether a success or failure was drawn, ensuring the urn's composition reflects prior outcomes. The hypergeometric distribution (Section 2.5) demonstrates the `urnTrials` construction.

#### 5.2.2.2 Sequential Trials until condition met

Some processes continue until a stopping condition is met, producing a variable (random) number of trials. This kind of repetition is modeled using the `iterate` kernel:

```
iterate : (S₁ →𝓔 (S₁ +ₛ S₂)) → (S₁ →𝓔 S₂) → (S₁ →𝓔 S₂)
iterate step loop = (loop ▿ₖ pureKernel) ∘ₖ step
```

The `step` kernel returns either a new state (left) or a final result (right). The `loop` kernel handles the continuation path, allowing the process to repeat with updated state.

Since truly infinite processes can't be directly computed, `unfoldₙ` approximates them using bounded recursion:

```
unfoldₙ : ℕ → S₁ →𝓔 S₂ → S₁ →𝓔 (S₁ +ₛ S₂) → S₁ →𝓔 S₂
unfoldₙ fuel tail = fix-approxₖ fuel tail ∘ iterate
```

The `fuel` parameter limits the number of steps. If the process hasn't terminated when `fuel` runs out, the `tail` kernel is used to resolve the remainder. The construction of the negative binomial distribution (Section 5.3.3) demonstrates use of `unfoldₙ`.

## 5.3 Summary

Statistical analysis often centers on summarizing experimental outcomes through counts, aggregations, and other summary statistics. Where earlier sections modeled how data are generated through choice and repetition, this section shows how applying summary operations to those processes produces classical probability distributions.

### 5.3.1 Multinomial

The multinomial construction demonstrates how summary statistics can emerge from the algebraic structure of repeated trials. Rather than defining the multinomial through its probability mass function, it arises naturally by folding over the convolution operation `∑[ n ]` to `n` independent `k`-way choices.

```
multinomial : (n : ℕ⁺) → Simplex k → ℰ (ℕₛ ^ k)
multinomial {k} n s = ∑[ n ] (const (kparts s ⟨$⟩ ≅⇒to tuple↔vec ∘→ unitVector k))
⟨$⟩ ≅⇒from tuple↔vec
  where open ConvolutionMonoid (vector-+-monoid k)
```

Each trial of `kparts s ⟨$⟩ unitVector k` produces a unit vector indicating the selected category. The convolution `∑[ n ]` then sums these indicator vectors across all trials, yielding the count vector directly. To demonstrate computation, consider 4 trials with uniform probability across 3 categories.

```
𝔼 = multinomial (4 #) (uniformSimplex (3 #))
```

The probability of observing counts $(1, 2, 1)$ is computed exactly as $\frac{4}{27}$, matching the probability mass function computation $\frac{4!}{1!\cdot 2!\cdot 1!} \times \left(\frac{1}{3}\right)^4$.

```
_ : PR (δℕ 1 ∧ δℕ 2 ∧ δℕ 1) ≟ + 4 / 27 ; _ = ✓
```

The following indicators do not sum to 4 thus PR is zero:

```
_ : PR (δℕ 0 ∧ δℕ 2 ∧ δℕ 1) ≟ 0ℚ ; _ = ✓
_ : PR (δℕ 2 ∧ δℕ 2 ∧ δℕ 1) ≟ 0ℚ ; _ = ✓
```

### 5.3.2 Binomial

The binomial distribution arises by convolving `n` independent Bernoulli trials, then mapping each trial outcome to its numerical value (1 for success, 0 for failure).

```
binomial : ℕ⁺ → P → ℰ ℕₛ
binomial n p = ∑[ n ] (const (bernoulli p ⟨$⟩ (1 ⟨▿⟩ 0)))
  where open ConvolutionMonoid ℕ.+-0-monoid
```

Consider 3 trials with success probability $\frac{1}{3}$:

```
𝔼 : ℰ ℕₛ
𝔼 = binomial (3 #) 1/[ₙ 3 ]
```

The probability of exactly 1 success computes to exactly $\frac{4}{9}$:

```
_ : PR (δℕ 1) ≟ + 4 / 9 ; _ = ✓
```

This corresponds to the theoretical binomial probability: $P(X = 1) = \binom{3}{1}c \cdot \left(\frac{1}{3}\right)^1 c \cdot \left(\frac{2}{3}\right)^2 = 3c \cdot \left(\frac{1}{3}\right)c \cdot \left(\frac{4}{9}\right) = \frac{4}{9} = 0.4444....$ For comparison, the R (R Core Team 2024) `dbinom` function for these binomial settings returns `0.44444444444444441977`. See Section 5.3.6 for further discussion of exact computations.

### 5.3.3 Negative Binomial

The negative binomial distribution models the number of failures observed before a fixed number of successes occurs in a sequence of independent Bernoulli trials. In this formalization, the process is expressed recursively by keeping track of the number of required successes still needed and the number of failures so far. Each Bernoulli trial either decrements the success count when a success is observed or increments the failure count when a failure is observed. When the success counter reaches one, the next success terminates the process and yields the final number of failures.

This update rule is encoded in the `stepNB` function, which advances the counters by a single Bernoulli draw:

```
stepNB : P → (ℕ⁺ₛ ×ₛ ℕₛ) →ℰ ((ℕ⁺ₛ ×ₛ ℕₛ) +ₛ ℕₛ)
stepNB p = ifₖ K (bernoulli p)
  then (idₖ ∘→ f)
  else (inj₁ₖ ∘ₖ (idₖ ⊗ₖ (idₖ ∘→ ℕsuc)))
  where
    f : (ℕ⁺ₛ ×ₛ ℕₛ) → ((ℕ⁺ₛ ×ₛ ℕₛ) +ₛ ℕₛ)
    to f (1 # , f)      = ⊎.inj₂ f
    to f (n +2 # , f)   = ⊎.inj₁ (n +1 # , f)
    -- cong proof of `f` elided
```

Repeatedly unfolding this step with a bounded "fuel" parameter yields an approximation to the negative binomial distribution:

```
negativeBinomialApprox : ℕ → ℕ⁺ → P → ℰ ℕₛ
negativeBinomialApprox fuel r p = to (unfoldₙ fuel proj₂ₖ (stepNB p)) (r , 0)
```

Here r is the number of required successes, p the success probability, and fuel the maximum number of iterations permitted.

For example, with three required successes and probability of success $p = \frac{1}{4}$:

```
𝔼 : ℰ ℕₛ
𝔼 = negativeBinomialApprox 100 (3 #) 1/[ₙ 4 ]
```

```
_ : PR (δℕ 1)  ≟ + 9 / 256  ; _ = ✓
```

A convergence proof showing that the approximation tends to the true distribution as fuel $\to \infty$ is deferred for future work.

### 5.3.4 Geometric

The geometric distribution is a special case of the negative binomial where the number of required successes is one. It models the number of trials required to observe the first success in a sequence of independent Bernoulli trials with success probability $p$.

This is obtained by reusing the stepNB construction with $r = 1$. The resulting distribution is then shifted by one so that its support is $\{1, 2, ...\}$, corresponding to the trial index of the first success:

```
geometric : P → ℰ ℕₛ
geometric p = negativeBinomialApprox 10 (1 #) p ⟨$⟩ ℕsuc
```

For example, with success probability $p = \frac{1}{4}$:

```
𝔼 : ℰ ℕₛ
𝔼 = geometric 1/[ₙ 4 ]
```

```
_ : PR (δℕ 4)  ≟ + 27 / 256  ; _ = ✓
_ : PR (𝟙≤ℕ 4) ≟ + 175 / 256 ; _ = ✓
```

As in the negative binomial case, the construction uses a finite fuel parameter to approximate the distribution.

### 5.3.5 Empirical

An empirical distribution places equal weight on a finite collection of observations. The constructor `empirical` builds an expectation space over `S` by taking the discrete uniform over `n` indices and mapping each index to its observed value. The result is a distribution that treats the supplied sample as the entire support, with mass $\frac{1}{n}$ on each point.

```
empirical : (n : ℕ⁺) → (Fin⁺ n → Ω S) → ℰ S
empirical n = (discreteUniform n ⟨$⟩_) ∘ mk→ᶠ
```

The `empiricalKernel` function lifts this construction to a kernel: given a function from the finite index space into `S`, it returns the corresponding empirical expectation. This is useful when the sample itself is a random output of an earlier stage in a model.

```
empiricalKernel : (n : ℕ⁺) → (Fin⁺ₛ n →ₛ S) →ℰ S
empiricalKernel n = record
  { to = empirical n ∘ to
  ; cong = empirical-cong
  }
```

The `sampleMean` function composes `empiricalKernel` with expectation, yielding a function from finite samples into the measurement space. It expresses the average as an estimator within the type-theoretic setting, making it available as a building block in estimation problems.

```
sampleMean : (n : ℕ⁺) → (S → ℳ.Mₛ) → (Fin⁺ₛ n →ₛ S) → ℳ.Mₛ
sampleMean n X = μ X ∘→ empiricalKernel n
```

#### 5.3.5.1 Scalar Empirical

In the scalar instance over rationals, expectations are exact. See Appendix C.2.1 for how any totally ordered field forms a `MAlgebra` of scalars.

```
𝔼 : ℰ ℚᵘₛ
𝔼 = empirical (5 #)
    λ { 0F → + 1  ℚᵘ./ 4
      ; 1F → + 11 ℚᵘ./ 8
      ; 2F → ℤ.- (+ 3) ℚᵘ./ 8
      ; 3F → ℤ.- (+ 90) ℚᵘ./ 17
      ; 4F → 0ℚᵘ
      }
```

```
_ : E id   ≟ ℤ.- (+ 55) / 68  ; _ = ✓
_ : Var id ≟ + 248313 / 46240 ; _ = ✓
```

The following example illustrates that conditioning works transparently on empirical distributions: define an indicator on `ℚᵘₛ` (e.g., $\leq 0$ vs $> 0$), construct $E[\mathrm{id} \mid IX]$, and evaluate the resulting conditional random variable at the two indicator values to obtain the corresponding conditional means.

```
X : Indicator ℚᵘₛ
X = 𝟙≤ℚᵘ 0ℚᵘ
```

```
XisSatisfiable : Satisfiable X
```

```
cond : E[ id | I X ]
cond = mkEY|I id (X , XisSatisfiable) ℚᵘ.≃-refl
```

```
-- ≤ 0
_ : to E[Y|X] 0ℚᵘ ≟ ℤ.- (+ 257) / 136 ; _ = ✓
-- > 0
_ : to E[Y|X] 1ℚᵘ ≟ (+ 13) / 16       ; _ = ✓
```

#### 5.3.5.2 Multi-dimensional Empirical

Anscombe's first dataset is encoded as an empirical distribution on $\mathbb{Q}^2$. See Appendix C.2.2 for how any totally ordered field forms a `MAlgebra` of vectors of scalars. The published decimals are entered as exact rationals, so the equalities that follow are literal rather than approximate.

```
𝔼 : ℰ Mₛ
𝔼 = empirical (11 #) λ
  {  0F → mk→ᶠ λ { 0F → + 10 ℚᵘ./ 1 ; 1F → + 201 ℚᵘ./ 25  }  -- 8.04
  ;  1F → mk→ᶠ λ { 0F → +  8 ℚᵘ./ 1 ; 1F → + 139 ℚᵘ./ 20  }  -- 6.95
  ;  2F → mk→ᶠ λ { 0F → + 13 ℚᵘ./ 1 ; 1F → + 379 ℚᵘ./ 50  }  -- 7.58
  ;  3F → mk→ᶠ λ { 0F → +  9 ℚᵘ./ 1 ; 1F → + 881 ℚᵘ./ 100 }  -- 8.81
  ;  4F → mk→ᶠ λ { 0F → + 11 ℚᵘ./ 1 ; 1F → + 833 ℚᵘ./ 100 }  -- 8.33
  ;  5F → mk→ᶠ λ { 0F → + 14 ℚᵘ./ 1 ; 1F → + 249 ℚᵘ./ 25  }  -- 9.96
  ;  6F → mk→ᶠ λ { 0F → +  6 ℚᵘ./ 1 ; 1F → + 181 ℚᵘ./ 25  }  -- 7.24
  ;  7F → mk→ᶠ λ { 0F → +  4 ℚᵘ./ 1 ; 1F → + 213 ℚᵘ./ 50  }  -- 4.26
  ;  8F → mk→ᶠ λ { 0F → + 12 ℚᵘ./ 1 ; 1F → + 271 ℚᵘ./ 25  }  -- 10.84
  ;  9F → mk→ᶠ λ { 0F → +  7 ℚᵘ./ 1 ; 1F → + 241 ℚᵘ./ 50  }  -- 4.82
  ; 10F → mk→ᶠ λ { 0F → +  5 ℚᵘ./ 1 ; 1F → + 142 ℚᵘ./ 25  }  -- 5.68
  }
```

Evaluation of `E id` returns the coordinate-wise means of the identity random variable.

```
_ : to (E id) 0F ≟ + 9 / 1       ; _ = ✓
_ : to (E id) 1F ≟ + 8251 / 1100 ; _ = ✓
```

To show the computation of the off-diagonal covariance, a reindexing random variable `X` swaps the two coordinates:

```
X : RV
to X f = mk→ᶠ λ {0F → to f 1F ; 1F → to f 0F }
```

```
_ : to (Cov id X) 0F ≟ + 5501 / 1100 ; _ = ✓
```

Across the examples, the same operators handle discrete choices, repeated trials, and empirical samples. Composing sampling processes with summary operations yields classical statistical quantities from both theoretical models and empirical data within the expectation framework.

### 5.3.6 Significance of Exact Computation

These examples highlight the connection between formal verification and exact numerical computation, demonstrating that the abstract operations correctly compute textbook formulas when instantiated with rational numbers. Using rational arithmetic ensures the calculations are mathematically exact, rather than approximations. The type-theoretic foundation guarantees distributions satisfy requisite axioms and probability calculations adhere to probability theory's formal requirements. This approach mitigates errors in statistical software such as:

- Implementation errors wherein code diverges from mathematical specifications;
- Numerical errors induced by floating-point approximations;
- Logical inconsistencies in the statistical formalizations.

Routines in existing statistical languages have decades of refinement to achieve their current precision and adhere to IEEE floating-point specifications, so their approximations may satisfy many practical requirements. In comparison, the approach developed in this paper bridges the divide between theoretical specification and computational implementation, ensuring their equivalence by construction rather than by post-hoc verification.

# 6 Beyond Expectations: Toward Correct by Construction Statistical Software

> Now, it is the contention of the intuitionists (or constructivists, I shall use these terms synonymously) that the basic mathematical notions, above all the notion of function, ought to be interpreted in such a way that the cleavage between mathematics, classical mathematics, that is, and programming that we are witnessing at present disappears.
>
> — from (Martin-Löf 1982)

This paper manifests Martin-Löf's vision by formalizing elementary statistical concepts in Agda. Building on expectation-based foundation of Whittle (2012), expectation is taken as the primitive notion from which probability, conditioning, and related constructs emerge. The expectation type $\mathcal{E}$ axiomatizes expectation spaces through properties that the operator `E` must satisfy: linearity, positivity, normalization, and compatibility with limits.

From this foundation arise five primitive operations (`pure`, `map`, `⟨*⟩`, `>>=`, `mix`) that compose into more complex distributions while preserving the expectation axioms. Convolution, conditioning, and other operations follow systematically. Implementations of standard distributions including Bernoulli, binomial, empirical, hypergeometric, and negative binomial showcase the computational content of the formalism. Instantiated with rational arithmetic, these examples perform exact computation without floating-point error and without divergence between code and mathematical specification.

Together, these results demonstrate how dependent type theory unifies logical rigor and executable correctness. This establishes a foundation for the development of verified statistical methods and points toward future extensions from elementary examples to full-scale applications where correctness and reproducibility are paramount.

## 6.1 Connections to Existing Work

### 6.1.1 Popular Statistical Programming Languages

While some languages popular in data science and statistics have type systems, such as Python's type hints and Julia's parametric types, these systems provide significantly less expressive power than dependent types. Python's type hints are optional annotations that help catch certain errors during development, but they do not offer formal guarantees or the ability to express relationships between values and types. Julia's parametric types enable more generic programming, yet they still lack value-dependent types that could encode properties such as “a probability distribution integrates to 1” or “an estimator is provably unbiased.” The approach presented here extends beyond these systems, offering a path to statistical software that is not only well-typed in the conventional sense but also mathematically verified through types that directly express statistical properties.

#### 6.1.2 Formal Methods

Dedicated venues such as the Workshop on Formal Statistics (Black 2019) show that interest in bridging statistics and formal verification is growing. Proof assistants such as Isabelle/HOL, Coq, and Lean have supported the development of measure theory and probability libraries, culminating in mechanically checked results such as the Central Limit Theorem (Avigad et al. 2017). The Isabelle probability library (Hasan and Tahar 2009; Mhamdi et al. 2013) and Lean’s mathlib (Doorn et al. 2020) provide extensive foundations for real analysis, measure theory, and probability. These efforts demonstrate that deep probabilistic results can be established within general-purpose proof assistants.

A parallel thread is probabilistic programming (Lew et al. 2023; Narayanan et al. 2016; Vákár et al. 2019), where languages such as Church, Anglican, Stan, and Pyro emphasize expressiveness and computational efficiency. These systems excel at practical inference, but typically do not offer the same formal guarantees as proof-assistant based approaches.

### 6.2 Implications for Statistical Practice

#### 6.2.1 Bridging Theory and Implementation

Traditional statistical software development separates mathematical specification from implementation. This work demonstrates an alternative where theory and code share the same formal foundation. The law of total expectation is simultaneously a proven theorem and an executable function. The binomial distribution is both a mathematical object satisfying probability axioms and a program computing exact probabilities.

This unification addresses the theory-implementation gap identified in the introduction. Software bugs arising from misinterpretation of specifications, coding errors in translating mathematics to programs, and drift between evolving theory and static implementations are prevented by construction. When theory changes, proofs must be updated; when proofs change, implementations automatically reflect those changes.

#### 6.2.2 Reproducibility and Verification

The “reproducibility crisis” in science has highlighted the critical importance of methodological transparency and computational reliability. Formal verification of statistical methods directly addresses several key aspects of this crisis. First, it eliminates implementation errors that can lead to irreproducible results. High-profile cases of irreproducibility due to software bugs, such as those in cancer research (Baggerly and Coombes 2009) and economics (Herndon et al. 2014), could have been prevented through formal verification of the computational methods used. Second, formalization requires explicit statement of assumptions, which are often implicit or underspecified in traditional statistical work. This explicitness reduces the risk of misapplication or misinterpretation of statistical methods, a common source of irreproducibility. Third, the integration of theory and implementation ensures that statistical software actually implements the methods it claims to, addressing the “implementation gap” that plagues many statistical packages. When theoretical guarantees are formally connected to computational implementations, users can have higher confidence in the validity of their analyses. Finally, the machine-checkable nature of formal proofs (Pollack 1998) provides a standard for correctness that can augment peer review. This is particularly important for complex statistical methods where errors might not be detected through traditional review processes.

#### 6.2.3 Teaching

Formal definitions force confronting conceptual issues that informal presentations may obscure. What does it mean for two outcomes to be equivalent? When is conditioning well-defined? The type system makes these questions explicit. The executable nature of proofs enables experimentation. Students can evaluate expressions, compute probabilities, and observe how operations compose. The Curry-Howard correspondence offers a conceptual framework for understanding statistical inference as constructing evidence. However, the learning curve is substantial. Whether formal methods improve statistical education remains an open question requiring empirical investigation.

### 6.3 Limitations and Challenges

Several limitations temper this work. The library covers elementary distributions and foundational operations. While the compositional nature of the primitives suggests that extensions may arise more naturally than in traditional approaches, demonstrating this for the breadth of applied statistical methods such as regression, hypothesis testing, survival analysis, and spatial statistics, remains to be explored.

Rational arithmetic ensures exactness but becomes prohibitively expensive for complex calculations. Denominators grow rapidly; operations slow correspondingly. Production statistical software accepts numerical error in exchange for speed. This work prioritizes correctness over performance, demonstrating what is possible in principle rather than what is practical at scale today.

The framework does not address hardware acceleration, distributed computation, or big data environments. Modern statistical computing relies on GPUs, TPUs, and distributed systems. Formal verification in these contexts introduces additional complexity: parallel algorithms contain subtle race conditions, distributed systems require coordination protocols, and approximations for tractability must be carefully managed.

### 6.4 Future Directions

#### 6.4.1 Theoretical Foundations

The expectation-based approach opens questions about relationships between foundational frameworks. Recent work on categorical probability theory (Fritz et al. 2023; Giry 1982; Perrone 2023) provides alternative axiomatizations where probability emerges from category-theoretic structures. Formal comparisons between expectation-based, measure-theoretic, and categorical foundations could clarify when each approach is advantageous and whether they are provably equivalent in restricted settings.

The relationship between constructive and classical probability deserves systematic investigation. Many classical results rely on non-constructive principles like excluded middle or choice. Reformulating these constructively may yield computational content or reveal which results genuinely require classical logic. This work intersects with synthetic probability theory and constructive measure theory, areas where foundational questions remain open.

#### 6.4.2 Practical Extensions

Several lines of development could extend the methodology. Collaborative library building is needed to cover common statistical methods beyond elementary probability. Such efforts should involve both domain experts and formal methods specialists, and be supported by educational resources that lower the barrier of entry for statisticians.

Improved numerical foundations are required. Integrating probabilistic error analysis and alternative numerical representations could balance correctness with practical tractability. Efficient compilation techniques will also be essential. Domain-specific languages and compilers that enable automatic parallelization and optimization for accelerators would bring formally verified code closer to practical data analysis. Scaling to big data environments poses a further challenge. Distributed and cloud-based computation introduces new correctness concerns. Yet correctness-first foundations may in fact enable stronger optimizations, by ruling out subtle synchronization errors and allowing approximation methods to be applied with rigorous error bounds.

### 6.5 Conclusion

Dependent type theory offers a path toward statistical software with correctness guarantees. This paper demonstrates feasibility through an expectation-based foundation formalized in Agda, establishing that mathematical rigor and computational execution can coexist. This work handles elementary distributions, proves foundational theorems with machine-checked proofs, and performs exact computation, showing that the theory-implementation gap can be closed by construction.

Significant challenges remain. Extending to comprehensive method coverage and achieving competitive performance require sustained effort. Yet the foundation is solid, the primitives compose predictably, and the examples show what becomes possible when proofs and programs are unified.

As proof assistants mature and formal methods become more accessible, verified statistical software transitions from aspiration to practical tool. The work presented here contributes to that trajectory: a foundation where statistical theory and programming are unified expressions of the same formal system, manifesting Martin-Löf's vision that the cleavage between mathematics and programming can disappear.

## Acknowledgements

This research was partially supported by NIH R01AI085073 and R01AI157758. The content is solely the responsibility of the authors and does not necessarily represent the official views of the NIH. Thanks to Steve Cole, Michael Hudgens, Michael Jetsupphasuk, and Brian Richardson for feedback on early drafts of this paper.

# A Introduction to Agda Syntax

This appendix provides an introduction to the Agda syntax used throughout the paper. Agda is a dependently typed programming language and proof assistant that allows us to express both mathematical definitions and their computational implementations in a unified framework.

## A.1 Basic Type Declarations and Definitions

In Agda, the type of an expression is declared using a colon (`:`), and its value is defined using an equals sign (=):

```
-- Declaring a natural number
n : ℕ         -- This declares n to be of type ℕ (natural numbers)
n = 0         -- This defines n as 0

-- Declaring a boolean
b : Bool      -- This declares b to be of type Bool (boolean values)
b = true      -- This defines b as true
```

Functions are declared with arrows (→) separating the parameter types from the return type:

```
-- A function that computes the maximum of two numbers
max : ℕ → ℕ → ℕ       -- Takes two natural numbers and returns a natural number
max zero    y       = y
max (suc x) zero    = suc x
max (suc x) (suc y) = suc (max x y) -- Recursive definition with pattern matching
```

## A.2 Lists and Vectors

Agda's standard library provides both non-dependent lists (`List`) and dependent vectors (`Vec`):

```
-- A list of natural numbers (with unknown length)
nums : List ℕ
nums = 1 ∷ 2 ∷ 3 ∷ []     -- The ∷ operator adds an element to a list
                          -- [] represents the empty list

-- A vector of natural numbers with length 3
vec : Vec ℕ 3
vec = 1 ∷ 2 ∷ 3 ∷ []      -- Same syntax, but type guarantees length is 3
```

## A.3 Universal Quantification and Implicit Arguments

The symbol ∀ (for all) introduces universal quantification and is often used with implicit arguments (in curly braces):

```
-- A polymorphic identity function
identity : ∀ {A : Type} → A → A     -- For any type A, takes an A and returns an A
identity x = x                      -- Simply returns its input

-- A function that creates a vector of a specific length
replicate : ∀ {A : Type} → (n : ℕ) → A → Vec A n
--          ^^^^^^^^^^^^   ^^^^^^^   ^   ^^^^^^^
--          For any type A  Take n    and a     Return a vector of A
```

```
--           (implicit arg)  a number  value     with exactly n elements
replicate zero     x = []
replicate (suc n) x = x ∷ replicate n x
```

The curly braces around `{A : Type}` indicate that this is an implicit argument. Agda will try to infer it from context rather than requiring it to be provided explicitly.

## A.4 Lambda Expressions

Lambda expressions (anonymous functions) are introduced using the lambda symbol (λ) followed by parameters and an arrow (→):

```
-- A lambda expression that squares its input
square : ℕ → ℕ
square = λ x → x * x

-- A function whose result is strictly positive
PositiveFunction : Type
PositiveFunction = Σ (ℕ → ℕ) λ f → ∀ x → f x > 0
--                 ^  ^        ^       ^
--                 |  |        |       Body of the lambda
--                 |  |        Parameter
--                 |  Type being quantified (here: functions between ℕ)
--                 Σ declares a dependent pair
```

## A.5 Dependent Function Types (Pi Types)

Dependent function types (Pi types) are represented using the arrow notation, where the return type can depend on the value of the parameter:

```
-- Safe lookup function (cannot produce out-of-bounds errors)
safe-lookup : ∀ {A : Type} {n : ℕ} → (v : Vec A n) → (i : Fin n) → A
--                              ^^^^^^^    ^^^^^^^^^^^    ^^^^^^^^^^    ^
--                              Length     Vector of A    Index < n     Return type
safe-lookup (x ∷ xs) zero     = x
safe-lookup (x ∷ xs) (suc i) = safe-lookup xs i
```

In the `safe-lookup` example, the type `Fin n` represents natural numbers less than `n`, ensuring that the index is always within bounds.

## A.6 Dependent Pair Types (Sigma Types)

Dependent pair types (Sigma types) are introduced using the Sigma symbol (Σ) followed by the type of the first component and a function that computes the type of the second component:

```
-- Type of positive natural numbers
PositiveNat : Type
PositiveNat = Σ ℕ λ n → n > 0
--            ^ ^  ^^^^   ^^^
--            | |  Function that  Returns the type
--            | Type of first     representing "n > 0"
--            | component
```

```
--            Sigma type constructor

-- Creating a value of type PositiveNat
two-positive : PositiveNat
two-positive = 2 , s≤s z≤n
--             ^   ^^^^^^^
--             |   Second component: proof that 2 > 0
--             |      s≤s is proof that (m≤n : m ≤ n) → suc m ≤ suc n
--             |      z≤n is proof that zero  ≤ n
--              First component: the number 2
```

The comma (,) separates the components of a pair.

## A.7 Pattern Matching and Case Analysis

Agda uses pattern matching to define functions by cases:

```
-- Boolean negation
not : Bool → Bool
not true = false   -- If input is true, return false
not false = true   -- If input is false, return true

-- Vector concatenation (showing dependent pattern matching)
_++_ : ∀ {A : Type} {m n : ℕ} → Vec A m → Vec A n → Vec A (m + n)
[] ++ ys = ys
(x ∷ xs) ++ ys = x ∷ (xs ++ ys)
```

## A.8 Inductive Types

Many types in Agda are defined inductively. For example, the following type could represent exposure in a clinical trial, where `Treated` is quantified by a natural number representing dosage.

```
data Exposure : Type where
  Untreated : Exposure
  Treated   : ℕ → Exposure
```

## A.9 Record Types

Records in Agda provide a way to define structures with named fields:

```
-- A simple record representing a point in 2D space
record Point : Type where
  constructor mkPoint   -- provide a data constuctor
  field
    x : ℕ
    y : ℕ

-- Constructing a term of type Point
p : Point
p = mkPoint 3 4
```

```
-- Extracting values from a Point
xofp : ℕ
xofp = Point.x p
```

Records are used extensively in this paper to represent mathematical structures with properties.

### A.10 Proofs and Evidence

In Agda, proofs are constructed as terms of proposition types. For example:

```
-- A proof that 2 + 2 = 4
two-plus-two : 2 + 2 ≡ 4
two-plus-two = refl   -- 'refl' is reflexivity - proof that a value equals itself
                      -- (this works because 2 + 2 computes to 4)
```

Throughout this paper, we use these syntactic constructs to formalize statistical concepts and prove their properties. While the syntax may initially seem unfamiliar, it provides a precise and machine-checkable language for expressing mathematical ideas with computational content. A full treatment of Agda or dependent type theory is beyond of the scope of this paper, and the interested reader is referred to several references (Agda Developers 2024; Bove et al. 2009; Wadler et al. 2022).

## B Sample Spaces

```
open import Relation.Binary public using (Setoid)
open import Relation.Binary.Indexed.Heterogeneous.Bundles public using
(IndexedSetoid)
open Setoid renaming (Carrier to Ω)
open import Function.Setoid public
  renaming
  ( const to K
  ; _∘_    to _∘→_
  )
open import Data.Product.Setoid public
  renaming
  ( ∏[_]⁺  to ∏[_]
  ; _⊗_    to _⊗ₛ_
  ; assocˡ to ×-assocˡ
  ; assocʳ to ×-assocʳ
  ; map₁   to ×-map₁
  ; map₂   to ×-map₂
  ; ×ₛ-distribˡ-⊎ₛ to ×ₛ-distribˡ-+ₛ
  )
open import Data.Product.Setoid.WithoutK public
open import Data.Sum.Setoid public
  renaming
  ( _⊎ₛ_     to _+ₛ_
  ; swap     to ⊎-swap
  ; ∐[_]⁺    to ∐[_]
  ; reduceₙ⁺ to reduceₙ
```

```
  ; _⊕_      to _⊕ₛ_
  ; assocˡ   to +-assocˡ
  ; assocʳ   to +-assocʳ
  ; map₂     to ⊎-map₂
  ; map₁     to ⊎-map₁
  ) hiding
  ( ∐[_]
  ; reduceₙ
  )
open import Data.Sum.Setoid.WithoutK public
```

```
SampleSpace = Setoid 0ℓ 0ℓ

ISampleSpace : SampleSpace → Type _
ISampleSpace I = IndexedSetoid (Ω I) 0ℓ 0ℓ
```

```
private
  variable
     a ℓa b ℓb : Level
     S S₁ S₂ : SampleSpace
```

## B.1 Sample Space Transformations

Sample space transformations play a central role in statistical theory, appearing in concepts such as invariance principles, symmetry, and change of variables. A transformation is simply a function that maps the sample space to itself, potentially altering the distribution of random variables.

```
_⟶_ : Setoid a ℓa → Setoid b ℓb → Type _
S₁ ⟶ S₂ = Func S₁ S₂
```

```
mk⟶ : (f : Ω S₁ → Ω S₂) → Congruent (_≈_ S₁) (_≈_ S₂) f → S₁ ⟶ S₂
mk⟶ f cong = record { to = f ; cong = cong }
```

```
_↠_ : Setoid a ℓa → Setoid b ℓb → Type _
S₁ ↠ S₂ = Surjection S₁ S₂
```

```
Endo : SampleSpace → Type
Endo S = S ⟶ S
```

## B.2 Instances

```
𝟙 : SampleSpace
𝟙 = ⊤.≡-setoid
```

```
𝟚 : SampleSpace
𝟚 = 𝟙 +ₛ 𝟙
```

```
Finₛ : (n : ℕ) → SampleSpace
Finₛ n = record
  { Carrier = Fin n
```

```
  ; _≈_ = _≡_
  ; isEquivalence = ≡.isEquivalence
  }
```

```
Fin⁺ₛ : (n : ℕ⁺) → SampleSpace
Fin⁺ₛ (n #) = record
  { Carrier = Fin n
  ; _≈_ = _≡_
  ; isEquivalence = ≡.isEquivalence
  }
```

```
ℕₛ : SampleSpace
ℕₛ = record { Carrier = ℕ ; _≈_ = _≡_ ; isEquivalence = ≡.isEquivalence }
```

```
ℕ⁺ₛ : SampleSpace
ℕ⁺ₛ = record { Carrier = ℕ⁺ ; _≈_ = _≡_ ; isEquivalence = ≡.isEquivalence }
```

```
ℕsuc : ℕₛ ⟶ ℕₛ
ℕsuc = mk⟶ ℕ.suc λ { ≡.refl → ≡.refl }
```

## B.3 Specialized Transformations

```
mk⟶ᶠ : ∀ {n} (f : Fin n → Ω S) → Finₛ n ⟶ S
mk⟶ᶠ {S = S} f = V.mkFunc S f
```

```
mkᴺ⟶ : (f : ℕ → Ω S) → ℕₛ ⟶ S
mkᴺ⟶ {S = S} f = mk⟶ f (F.≡⇒Congruent {A = S} {D = ℕ} f)
```

## B.4 Isomorphisms

```
_≃ₛ_ : SampleSpace → SampleSpace → Type
S₁ ≃ₛ S₂ = S₁ ≃ S₂
```

```
≅⇒to : S₁ ≃ₛ S₂ → S₁ ⟶ S₂
≅⇒to = Inv.toFunction
```

```
≅⇒from : S₁ ≃ₛ S₂ → S₂ ⟶ S₁
≅⇒from = Inv.fromFunction
```

```
Permutation : SampleSpace → Type
Permutation S = S ≃ₛ S
```

```
proj-iso : ∀ {n⁺@(n #) : ℕ⁺} → ∏[ n +1 # ] (const S) ≃ₛ (S ×ₛ ∏[ n⁺ ] (const S))
```

The Setoid of sample spaces.

```
SampleSpaceₛ : Setoid (ℓsuc 0ℓ) 0ℓ
SampleSpaceₛ = record
  { Carrier = SampleSpace
  ; _≈_ = _≃ₛ_
  ; isEquivalence = Inv.isEquivalence
```

```
  }
```

```
𝓢-map : SampleSpace → Type₁
𝓢-map S = Func S SampleSpaceₛ
```

## B.5 Etc

```
inj₁⁺ : ∀ {n} → (2 ^ ((n +1) #)) → (2 ^ ((n +2) #))
inj₁⁺ = mk⟶ (⊎.inj₁ tt ,_) (lift ≡.refl ,_)

inj₂⁺ : ∀ {n} → (2 ^ ((n +1) #)) → (2 ^ ((n +2) #))
inj₂⁺ = mk⟶ (⊎.inj₂ tt ,_) (lift ≡.refl ,_)
```

```
isOne : ℕ⁺ₛ ⟶ 𝟚
to isOne (1 #) = ⊎.inj₁ tt
to isOne (_ +2 #) = ⊎.inj₂ tt
cong isOne {1 #} {1 #} ≡.refl = lift ≡.refl
cong isOne {_ +2 #} {_ +2 #} ≡.refl = lift ≡.refl
```

```
if_Then_Else : (S ⟶ 𝟚) → (S ⟶ S₁) → (S ⟶ S₂) → S ⟶ (S₁ +ₛ S₂)
if test Then f Else g = (f ⊕ₛ g) ∘⟶ branch test
```

```
rep-prim : ∀ {n} → Ω S → (Ω (S ^ n))
rep-prim {n = 1 #} x = x
rep-prim {n = n +2 #} x = x , rep-prim {n = n +1 #} x

rep-cong : ∀ {n} → Congruent (_≈_ S) (_≈_ (S ^ n)) (rep-prim {n = n})
rep-cong {n = 1 #} x = x
rep-cong {n = n +2 #} x = x , rep-cong {n = n +1 #} x

rep : ∀ n → S ⟶ (S ^ n)
rep n = mk⟶ (rep-prim {n = n}) (rep-cong {n = n})
```

```
replicate : ∀ n → S ⟶ (Fin⁺ₛ n →ₛ S)
to (replicate n) = K
cong (replicate n) x = x
```

```
∐𝟙→Fin-prim : ∀ {n} → (Ω (∐[ n ] (λ _ → 𝟙))) → (Ω (Fin⁺ₛ n))
∐𝟙→Fin-prim {1 #} _ = zero
∐𝟙→Fin-prim {n +2 #} (⊎.inj₁ _) = zero
∐𝟙→Fin-prim {n +2 #} (⊎.inj₂ i) = suc (∐𝟙→Fin-prim {n +1 #} i)

∐𝟙→Fin-cong : ∀ {n} → Congruent (_≈_ (∐[ n ] (const 𝟙))) _≡_ (∐𝟙→Fin-prim {n})
∐𝟙→Fin-cong {1 #} _ = ≡.refl
∐𝟙→Fin-cong {n +2 #} {⊎.inj₁ _} {⊎.inj₁ _} _ = ≡.refl
∐𝟙→Fin-cong {n +2 #} {⊎.inj₂ _} {⊎.inj₂ _} (lift x≡y) = ≡.cong suc (∐𝟙→Fin-cong {n
+1 #} x≡y)

∐𝟙→Fin : ∀ {n} → (∐[ n ] (λ _ → 𝟙)) ⟶ (Fin⁺ₛ n)
```

```
∐𝟙→Fin = mk→ ∐𝟙→Fin-prim ∐𝟙→Fin-cong
```

```
countSuccesses : ∀ n → ∏[ n ] (const 2) → Fin⁺ₛ ((ℕ⁺.# n) +1 #)
countSuccesses n = mk→ (countSuccessesPrim {n}) (countSuccesses-cong {n})
```

```
countUptoK : (K : ℕ) (n : ℕ⁺) → (2 ^ n) → Finₛ (K +1)
countUptoK K n = mk→ (countUptoKprim K n) (countUptoK-cong K n)
```

# C Type of container for measurement values

The `MAlgebra` structure extends the familiar notion of a vector space to provide a suitable setting for measurement values in statistical applications. It is a partially ordered module over a totally ordered scalar field, enriched with a trace and a seminorm so that statistical operations have clear algebraic and analytic meaning.

Concretely, an `MAlgebra` equips the space of measurement values with:

- **Base algebra**:

a `PoUnitalAlgebra` (partially ordered unital algebra), giving linear and multiplicative structure together with a partial order on the module.

- **Trace**:

a functional mapping measurement values back to scalars in the underlying field, supporting the definition of probability.

- **Semi-norm**:

an operation measuring the "size" or "magnitude" of values, used to express inequalities and convergence.

Compatibility conditions ensure that the order, trace, and seminorm interact coherently, so algebraic transformations preserve statistical meaning. This makes `MAlgebra` a foundation for expectation spaces ($\mathcal{E}$), where functions may take values not just in scalars but also in richer algebraic structures such as vectors or matrices.

### Remarks

- On ordering:

The scalar field is assumed totally ordered: for any two scalars $a$, $b$, either $a \leq b$ or $b \leq a$.

- This holds for rationals constructively, and for reals in the classical setting.
- In purely constructive analysis, the reals cannot be shown to satisfy this property; they form only a partial order unless one assumes additional principles (e.g. excluded middle).
- The module itself is only partially ordered, with the order compatible with its linear structure, trace, and seminorm.
- The present definition is intentionally minimal, capturing only the algebraic structure required for the examples in this paper. It can be specialized or extended with stronger structure (norms, topologies, etc.) depending on application.

## C.1 `M-Algebra`

```
record MAlgebra (ℛ : ToField 0ℓ 0ℓ 0ℓ 0ℓ) : Type₁ where


  private
    open module R = ToField ℛ
```

```
      using
      (0# ; 1# ; Carrier; _≤_; _≈_; _+_
      ; _*_
      ; _-_
      ; -_
      ; +-congˡ
      ; |_|
      ; 1/
      ; 𝕟 )

private
  open module FieldProps = ToFieldProps ℛ
     using (1/[𝕟_])

infixl 6 _-ᴹ_

field
  algebra : PoUnitalAlgebra R.poCommutativeRing 0ℓ 0ℓ 0ℓ

open PoUnitalAlgebra algebra public
open Trace R.poCommutativeSemiring
open Norm R.poCommutativeSemiring (ToFieldProps.toFieldHasValuation ℛ)
open NS⇒S _≈ᴹ_ _≤ᴹ_ public
  renaming
    ( _<_      to _<ᴹ_
    ; <-irrefl to <ᴹ-irrefl
    ; <⇒≤      to <ᴹ⇒≤ᴹ
    )

field
  <ᴹ-nonTrivial : 0ᴹ <ᴹ 1ᴹ

--- Trace ------------
field
  trace : Traced (PoUnitalLeftAlgebra.poUnitalLeftSemialgebra poUnitalLeftAlgebra)

open Traced trace public renaming (
      cong         to Tr-cong
    ; +-homo       to Tr-+-homo
    ; *ₗ-homo      to Tr-*ₗ-homo
    ; isPositive   to Tr-positive
    ; isUnital     to Tr-unital
    ; f0≈0         to Tr0≈0
   )
   using (Tr-cyclic; Tr; Tr-functional)
```

```
  TrFunc : Func ≈ᴹ-setoid R.setoid
  TrFunc = record { to = Tr ; cong = Tr-cong }

  --- Seminorm -----------
  field
    ‖_‖ᴹ : Carrierᴹ → R.Carrier
    hasSemiNorm : IsSeminorm leftSemimodule ‖_‖ᴹ

  open IsSeminorm hasSemiNorm public renaming
    ( cong        to ‖·‖-cong
    ; nonNegative to ‖·‖-nonNegative
    ; |•|-homo    to ‖·‖-*ₗ-homo
    ; triangle    to ‖·‖-triangle
    )

  ‖·‖ᶠ : Func ≈ᴹ-setoid R.setoid
  ‖·‖ᶠ = record { to = ‖_‖ᴹ ; cong = ‖·‖-cong }
```

### C.1.1 Sample Spaces

```
  Mₛ : SampleSpace
  Mₛ = ≈ᴹ-setoid
```

## C.2 Constructions

The following examples illustrate two concrete `MAlgebra` constructions used in this paper: one for scalar measurement values, and one for finite-dimensional vectors.

### C.2.1 TensorUnit

The simplest instance of an `MAlgebra` arises from the scalar field itself. The `mkScalarAlgebra` function builds the unit measurement space, in which every value is a scalar and the trace and seminorm are the identity and absolute-value operations (respectively) of the field. This serves as the tensor unit for higher-dimensional constructions: expectations in this space correspond to ordinary numeric expectations.

```
mkScalarAlgebra : (F : ToField 0ℓ 0ℓ 0ℓ 0ℓ) → MAlgebra F
mkScalarAlgebra F = record
  { algebra = TU.mkPoUnitalAlgebra
  ; <ᴹ-nonTrivial = nonTrivial , #⇒≉ (#-sym 1#0)
  ; trace = TraceUnit.mkTraced poCommutativeSemiring
  ; hasSemiNorm = NormUnit.|-|-isSeminorm poCommutativeSemiring toFieldHasValuation
  }
  where open ToField F
        open ToFieldProps F
```

### C.2.2 Vectors

The `mkVectorAlgebra` function lifts the scalar `MAlgebra` to finite-dimensional vector spaces. Given a field `F` and a dimension `n`, it constructs an `MAlgebra` whose carrier is `Fin⁺ n → F` (vectors). The partial order, trace, and seminorm are all defined pointwise, ensuring compatibility with the scalar algebra. The resulting structure supports expectations of vector-valued random variables, with the trace acting as a summation over components and the seminorm derived from the scalar seminorm coordinatewise.

```
mkVectorAlgebra : (F : ToField 0ℓ 0ℓ 0ℓ 0ℓ) → (n : ℕ⁺) → MAlgebra F
mkVectorAlgebra F n = record
  { algebra = P.mkPoUnitalAlgebra (Fin⁺ₛ n) TU.mkPoUnitalAlgebra
  ; <ᴹ-nonTrivial = nonTrivial , λ x → #⇒≉ (#-sym 1#0) (x { first {n} })
  ; trace = TraceVector.mkTracedFunc F (ℕ⁺.# n)
  ; hasSemiNorm = NormVector.mkIsSeminorm
       poCommutativeSemiring toFieldHasValuation M.leftSemimodule
       M.‖·‖ᶠ M.hasSemiNorm (ℕ⁺.# n)
  }
  where open ToField F
        open ToFieldProps F
        module M = MAlgebra (mkScalarAlgebra F)
```

# D Functions valued in a MAlgebra

```
open module RV = PoUnitalAlgebra
  (P.mkPoUnitalAlgebra S (MAlgebra.algebra 𝓜)) public
  using ()
  renaming (
      _+ᴹ_ to _+′_
    ; _*ᴹ_ to _*′_
    ; -ᴹ_  to -′_
    ; _*ₗ_ to _*ₗ′_
    ; _*ᵣ_ to _*ᵣ′_
    ; _≤ᴹ_ to _≤′_
    ; _≈ᴹ_ to _≈′_
    ; 0ᴹ   to 0ᴹ′
    ; 1ᴹ   to 1ᴹ′
    ; Carrierᴹ to RV
    ; ≈ᴹ-setoid to RVₛ
  )

RVring : PoRing 0ℓ 0ℓ 0ℓ
RVring = Pw.mkPoRing S poRingᴹ
```

## D.1 Additional Operations

```
_-′_ : RV → RV → RV
X -′ Y = X +′ (-′ Y)

‖_‖′ : RV → RV
‖ X ‖′ = mk↝ (λ ω → ‖ to X ω ‖ᴹ *ₗ 1ᴹ) λ ω≈ω′ → *ₗ-congʳ (‖·‖-cong (cong X ω≈ω′))

_²′ : RV → RV
X ²′ = X *′ X
```

## D.2 Special RVs

### D.2.1 Indicator RVs

```
open module 𝕀 = Indicator (PoRing.semiring poRingᴹ)
  using (Indicator; IsIndicator; mk)
  renaming
    (⟦_⟧ to I)
  public
open 𝕀.Indicator public renaming (I to toRV)

open import Function.Indicator.Setoid.Properties
  (PoRing.semiring poRingᴹ) public
open import Function.Indicator.Setoid.Base
  (PoRing.ring poRingᴹ) public
open import Function.Indicator.Setoid.Properties.PoSemiring
  (PoRing.poSemiring poRingᴹ) public
open import Function.Indicator.Setoid.Properties.Ring
  (PoRing.ring poRingᴹ) public
```

### D.2.2 Sequences

```
Monotonic↗ = Seq.Monotonic↗ _≈′_ _≤′_

_ConvergesPointwiseTo_ : Sequence RV → RV → Type
Xₙ ConvergesPointwiseTo X = ∀ (ω : Ω) → (λ i → to (Xₙ i) ω) ConvergesTo to X ω
```

### D.2.3 NonNegative RVs

A type for non-negative random variables is defined as:

```
record RV⁰⁺ : Type where
  constructor mk
  field
    X : RV
    nonNeg : 0ᴹ′ ≤′ X
```

## D.3 Constructing RVs

### D.3.1 Inequality indicator RVs

```
X≤?xpre : (Ω → Carrierᴹ) → R → (Ω → Carrierᴹ)
X≤?xpre X r ω with total ‖ X ω ‖ᴹ r
... | (⊎.inj₁ X≤x) = 1ᴹ
... | (⊎.inj₂ x≤X) = 0ᴹ

X≤?x-cong : ∀ ((mk X _) : RegularRV) (r : R) → Congruent _≈_ _≈ᴹ_ (X≤?xpre (to X) r)
X≤?x-cong (mk X safe) r {ω} {ω′} ω≈ω′ with total ‖ to X ω ‖ᴹ r | total ‖ to X ω′ ‖ᴹ r
... | (⊎.inj₁ Xω≤x) | (⊎.inj₁ Xω′≤x) = ≈ᴹ-refl
... | (⊎.inj₂ x≤Xω) | (⊎.inj₂ x≤Xω′) = ≈ᴹ-refl
... | (⊎.inj₁ Xω≤x) | (⊎.inj₂ x≤Xω′) =
```

```
 ⊥-elim (
    ℛ.irrefl
      (ℛ.antisym Xω≤x (ℛ.≤-respʳ-≈ (‖·‖-cong (Eq.sym (cong X  ω≈ω′ ))) x≤Xω′))
      (safe r ω≈ω′ Xω≤x x≤Xω′))
... | (⊎.inj₂ x≤Xω) | (⊎.inj₁ Xω′≤x) =
  ⊥-elim
     (ℛ.irrefl
       (ℛ.antisym Xω′≤x (ℛ.≤-respʳ-≈ (‖·‖-cong (cong X ω≈ω′)) x≤Xω))
       (safe r (sym ω≈ω′) Xω′≤x x≤Xω))

X≤?x : RegularRV → R → S → Mₛ
X≤?x R@(mk X safe) x = mk→ (X≤?xpre (to X) x) (X≤?x-cong R x)

X≤?xind : (X : RegularRV) → (x : R) → IsIndicator (X≤?x X x)
X≤?xind (mk X _) x ω with total ‖ to X ω ‖ᴹ x
... | (⊎.inj₁ X≤x) = ⊎.inj₁ Eq.refl
... | (⊎.inj₂ x≤X) = ⊎.inj₂ Eq.refl

_≤?_ : RegularRV → R → Indicator S
X ≤? x = Indicator.mk (X≤?x X x) (X≤?xind X x)

X≥?xpre : (Ω → Carrierᴹ) → R → Ω → Carrierᴹ
X≥?xpre X x ω with total ‖ X ω ‖ᴹ x
... | (⊎.inj₁ X≤x) = 0ᴹ
... | (⊎.inj₂ x≤X) = 1ᴹ

X≥?x-cong : ∀ ((mk X _) : RegularRV) (r : R) → Congruent _≈_ _≈ᴹ_ (X≥?xpre (to X) r)
X≥?x-cong (mk X safe) r {ω} {ω′} ω≈ω′ with total ‖ to X ω ‖ᴹ r | total ‖ to X ω′ ‖ᴹ r
... | (⊎.inj₁ Xω≤x) | (⊎.inj₁ Xω′≤x) = ≈ᴹ-refl
... | (⊎.inj₂ x≤Xω) | (⊎.inj₂ x≤Xω′) = ≈ᴹ-refl
... | (⊎.inj₁ Xω≤x) | (⊎.inj₂ x≤Xω′) =
 ⊥-elim (
    ℛ.irrefl
      (ℛ.antisym Xω≤x (ℛ.≤-respʳ-≈ (‖·‖-cong (Eq.sym (cong X ω≈ω′))) x≤Xω′))
      (safe r ω≈ω′ Xω≤x x≤Xω′))
... | (⊎.inj₂ x≤Xω) | (⊎.inj₁ Xω′≤x) =
  ⊥-elim
     (ℛ.irrefl
       (ℛ.antisym Xω′≤x (ℛ.≤-respʳ-≈ (‖·‖-cong (cong X ω≈ω′)) x≤Xω))
       (safe r (sym ω≈ω′) Xω′≤x x≤Xω))

X≥?x : RegularRV → R → S → Mₛ
X≥?x R@(mk X safe) x = mk→ (X≥?xpre (to X) x) (X≥?x-cong R x)

X≥?xind : (X : RegularRV ) → (x : R) → IsIndicator (X≥?x X x)
X≥?xind X x ω  with total ‖ to (rv X) ω ‖ᴹ x
```

```
... | (⊎.inj₁ X≤x) = ⊎.inj₂ Eq.refl
... | (⊎.inj₂ x≤X) = ⊎.inj₁ Eq.refl

_≥?_ : RegularRV → R → Indicator S
X ≥? x = Indicator.mk (X≥?x X x) (X≥?xind X x)
```

## D.4 Properties

The following lemma states scaling the indicator random variable `X≥?x` by `r` is less than or equal to the normed random variable `X`. This property is used to prove Markov's inequality.

```
r*‖X≥r‖≤‖X‖ : ∀ (r : R) → (X : RegularRV⁰⁺)
  → r *ₗ′ X≥?x (regular X) r ≤′ ‖ rv⁰⁺ X ‖′
r*‖X≥r‖≤‖X‖ r X {ω} with total ‖ to (rv⁰⁺ X) ω ‖ᴹ r
... | inj₁ X≤r = begin
  r *ₗ 0ᴹ          ≈⟨ *ₗ-zeroʳ r ⟩
  0ᴹ               ≈⟨ *ₗ-zeroʳ _ ⟨
  ‖ to (rv⁰⁺ X) ω ‖ᴹ *ₗ 0ᴹ ≤⟨ *ₗ-compatₗ ‖·‖-nonNegative (<ᴹ⇒≤ᴹ <ᴹ-nonTrivial) ⟩
  ‖ to (rv⁰⁺ X) ω ‖ᴹ *ₗ 1ᴹ ∎
  where open <-Reasoning ≤ᴹ-poset
... | inj₂ r≤X = begin
  r *ₗ 1ᴹ          ≤⟨ *ₗ-compatᵣ r≤X (<ᴹ⇒≤ᴹ <ᴹ-nonTrivial) ⟩
  ‖ to (rv⁰⁺ X) ω ‖ᴹ *ₗ 1ᴹ ∎
  where open <-Reasoning ≤ᴹ-poset
```

### D.4.1 Properties of Indicator RVs

```
module _ (I : Indicator S) where

  private
    X  = toRV I
    Xᶜ = toRV (C I)

  X≈1⇒Xᶜ≈0 : ∀ {ω} → to X ω ≈ᴹ 1ᴹ → 1ᴹ -ᴹ to X ω ≈ᴹ 0ᴹ
  X≈1⇒Xᶜ≈0 Xω≈1 = Eq.trans (+ᴹ-congˡ (-ᴹ‿cong Xω≈1)) (-ᴹ‿inverseʳ 1ᴹ)

  X≈0⇒Xᶜ≈1 : ∀ {ω} → to X ω ≈ᴹ 0ᴹ → to Xᶜ ω ≈ᴹ 1ᴹ
  X≈0⇒Xᶜ≈1 Xω≈0 = Eq.trans
    (+ᴹ-congˡ (Eq.trans (-ᴹ‿cong Xω≈0) -0ᴹ≈0ᴹ))
    (+ᴹ-identityʳ _)

  X*Xᶜ≈0 : X *′ Xᶜ ≈′ 0ᴹ′
  X*Xᶜ≈0 {ω} with isIndicator I ω
  ... | inj₁ Xω≈1 = Eq.trans (*ᴹ-congˡ (X≈1⇒Xᶜ≈0 Xω≈1)) (0ᴹ-zeroʳ _)
  ... | inj₂ Xω≈0 = Eq.trans (*ᴹ-congʳ Xω≈0) (0ᴹ-zeroˡ _)

  X*ᵣTrX≈X⋆ᵣTr1 : ∀ {ω} → to X ω *ᵣ Tr (to X ω) ≈ᴹ to X ω *ᵣ Tr 1ᴹ
  X*ᵣTrX≈X⋆ᵣTr1  {ω} with isIndicator I ω
```

```
  ... | inj₁ Xω≈1 = *ᵣ-congˡ (Tr-cong Xω≈1)
  ... | inj₂ Xω≈0 = 0ᴹ-cancel-*ᵣ Xω≈0

  X*ᵣTrXᶜ≈0  : ∀ {ω} → to X ω *ᵣ Tr (to Xᶜ ω) ≈ᴹ 0ᴹ
  X*ᵣTrXᶜ≈0 {ω} with isIndicator I ω
  ... | inj₁ Xω≈1 = begin
      to X ω *ᵣ Tr (to Xᶜ ω) ≈⟨ *ᵣ-congˡ (Tr-cong (X≈1⇒Xᶜ≈0 Xω≈1)) ⟩
      to X ω *ᵣ Tr 0ᴹ         ≈⟨ *ᵣ-congˡ Tr0≈0 ⟩
      to X ω *ᵣ 0#            ≈⟨ *ᵣ-zeroʳ _ ⟩
      0ᴹ ∎
      where open ≈-Reasoning ≈ᴹ-setoid
  ... | inj₂ Xω≈0 = begin
      to X ω *ᵣ Tr (to Xᶜ ω) ≈⟨ *ᵣ-congʳ Xω≈0 ⟩
      0ᴹ  *ᵣ Tr (to Xᶜ ω)     ≈⟨ *ᵣ-zeroˡ _ ⟩
      0ᴹ
      ∎
      where open ≈-Reasoning ≈ᴹ-setoid

module _ (X : Indicator S) (f : Mₛ → Mₛ) where

  X∘fX≈X*f1 : toRV X *′ f ∘→ (toRV X) ≈′ toRV X *′ K (to f 1ᴹ)
  X∘fX≈X*f1 {ω} with isIndicator X ω
  ... | inj₁ Xω≈1 = *ᴹ-congˡ (cong f Xω≈1)
  ... | inj₂ Xω≈0 = 0ᴹ-cancelˡ Xω≈0

  X∘fCX≈X*f0 : toRV X *′ f ∘→ (toRV (C X)) ≈′ toRV X *′ K (to f 0ᴹ)
  X∘fCX≈X*f0 {ω} with isIndicator X ω
  ... | inj₂ Xω≈0 = 0ᴹ-cancelˡ Xω≈0
  ... | inj₁ Xω≈1 = *ᴹ-congˡ (cong f lemma)
      where lemma : 1ᴹ -ᴹ to (toRV X) ω ≈ᴹ 0ᴹ
            lemma = Eq.trans (+ᴹ-congˡ (-ᴹ‿cong Xω≈1)) (-ᴹ‿inverseʳ 1ᴹ)

module _ (I : Indicator S) (f : Mₛ → Mₛ) where

  private
    X  = toRV I
    Xᶜ = toRV (C I)

  f∘X≈f1X+f0Xᶜ : f ∘→ X ≈′ K (to f 1ᴹ) *′ X +′ K (to f 0ᴹ) *′ Xᶜ
  f∘X≈f1X+f0Xᶜ  {ω} with isIndicator I ω
  ... | inj₁ Xω≈1 = begin
      to f (to X ω)                        ≈⟨ +ᴹ-identityʳ _ ⟨
      to f (to X ω) +ᴹ 0ᴹ                  ≈⟨ +ᴹ-cong (cong f Xω≈1) (Eq.sym (0ᴹ-
      zeroʳ _)) ⟩
      to f 1ᴹ +ᴹ to f 0ᴹ *ᴹ 0ᴹ             ≈⟨ +ᴹ-cong (*ᴹ-identityʳ _) (*ᴹ-congˡ
      (X≈1⇒Xᶜ≈0 I Xω≈1)) ⟨
```

```
    to f 1ᴹ *ᴹ 1ᴹ +ᴹ to f 0ᴹ *ᴹ to Xᶜ ω       ≈⟨ +ᴹ-congʳ (*ᴹ-congˡ (Eq.sym
    Xω≈1)) ⟩
    to f 1ᴹ *ᴹ to X ω +ᴹ to f 0ᴹ *ᴹ to Xᶜ ω
    ∎
    where open ≈-Reasoning ≈ᴹ-setoid
... | inj₂ Xω≈0 = begin
    to f (to X ω)                             ≈⟨ +ᴹ-identityˡ _ ⟨
    0ᴹ +ᴹ to f (to X ω)                       ≈⟨ +ᴹ-cong (Eq.sym (0ᴹ-zeroʳ _))
    (cong f Xω≈0) ⟩
    to f 1ᴹ *ᴹ 0ᴹ +ᴹ to f 0ᴹ                  ≈⟨ +ᴹ-cong (*ᴹ-congˡ Xω≈0) (*ᴹ-
    identityʳ _) ⟨
    to f 1ᴹ *ᴹ to X ω +ᴹ to f 0ᴹ *ᴹ 1ᴹ        ≈⟨ +ᴹ-congˡ (*ᴹ-congˡ lemma) ⟩
    to f 1ᴹ *ᴹ to X ω +ᴹ to f 0ᴹ *ᴹ to Xᶜ ω
    ∎
    where open ≈-Reasoning ≈ᴹ-setoid
          lemma : 1ᴹ ≈ᴹ  to Xᶜ ω
          lemma = Eq.trans (Eq.sym (+ᴹ-identityʳ _)) (+ᴹ-congˡ (Eq.trans (Eq.sym
          -0ᴹ≈0ᴹ) (-ᴹ‿cong (Eq.sym Xω≈0))))
```